\documentclass[a4paper,fleqn,usenatbib]{mnras}

\usepackage{amssymb}
\usepackage{amsmath}
\usepackage{txfonts}

\usepackage[T1]{fontenc}
\usepackage{ae,aecompl}

\usepackage{graphicx}
\usepackage{caption}
\usepackage{url}
\usepackage{multirow}
\usepackage{subfig}
\usepackage{array}
\usepackage{amssymb}
\usepackage{xtab,booktabs}
\usepackage{mathtools}
\usepackage{epstopdf}
\usepackage{float}
\usepackage{tabularx}
\usepackage[table]{xcolor}
\definecolor{RobGreen}{RGB}{146,232,156}
\definecolor{RobYellow}{RGB}{232,232,146}
\definecolor{RobRed}{RGB}{232,146,146}
\usepackage{algorithm}
\usepackage[noend]{algpseudocode}

\title[Fifty Years of Pulsar Candidate Selection]{Fifty Years of Pulsar Candidate Selection: From simple filters to a new principled real-time classification approach}

\author[R. J. Lyon et al.] {R. J.~Lyon $^1$\thanks{E-mail: robert.lyon@cs.man.ac.uk}, B. W.~Stappers $^2$, S.~Cooper $^2$, J. M.~Brooke $^1$, J. D.~Knowles $^{1,3}$  \newauthor 
\\ $^1$School of Computer Science, The University of Manchester, Manchester M13 9PL
\\ $^2$Jodrell Bank Centre for Astrophysics, School of Physics and Astronomy, The University of Manchester, Manchester M13 9PL
\\ $^3$School of Computer Science, University of Birmingham, Edgbaston, Birmingham B15 2TT}

\date{Accepted XXX. Received YYY; in original form ZZZ}

\pubyear{2015}

\begin{document}
\label{firstpage}
\pagerange{\pageref{firstpage}--\pageref{lastpage}}
\maketitle

\begin{abstract}
Improving survey specifications are causing an exponential rise in pulsar candidate numbers and data volumes. We study the candidate filters used to mitigate these problems during the past fifty years. We find that some existing methods such as applying constraints on the total number of candidates collected per observation, may have detrimental effects on the success of pulsar searches. Those methods immune to such effects are found to be ill-equipped to deal with the problems associated with increasing data volumes and candidate numbers, motivating the development of new approaches. We therefore present a new method designed for on-line operation. It selects promising candidates using a purpose-built tree-based machine learning classifier, the Gaussian Hellinger Very Fast Decision Tree (GH-VFDT), and a new set of features for describing candidates. The features have been chosen so as to i) maximise the separation between candidates arising from noise and those of probable astrophysical origin, and ii) be as survey-independent as possible. Using these features our new approach can process millions of candidates in seconds ($\sim$1 million every 15 seconds), with high levels of pulsar recall (90\%+). This technique is therefore applicable to the large volumes of data expected to be produced by the Square Kilometre Array (SKA). Use of this approach has assisted in the discovery of 20 new pulsars in data obtained during the LOFAR Tied-Array All-Sky Survey (LOTAAS).
\end{abstract}




\section{Introduction}
The search techniques used to isolate the radio emission of pulsars, are designed to find periodic broadband signals exhibiting signs of dispersion caused by travel through the interstellar medium. Signals meeting these criteria are recorded as a collection of diagnostic plots and summary statistics, in preparation for analysis. Together these plots and statistics are referred to as a pulsar `candidate', a  possible detection of a new pulsar. Each candidate must be inspected by either an automated method, or a human expert, to determine their authenticity. Those of likely pulsar origin are highlighted for further analysis, and possibly allocated telescope time for confirmation observations. The remainder are typically ignored. The process of deciding which candidates are worthwhile investigating has become known as candidate `selection'. It is an important step in the search for pulsars since it allows telescope time to be prioritised upon those detections likely to yield a discovery. Until recently (early 2000's) candidate selection was a predominately manual task. However advances in telescope receiver design, and the capabilities of supporting computational infrastructures, significantly increased the number of candidates produced by modern pulsar surveys \citep{Stovall:2013:dl}. Manual approaches therefore became impractical, introducing what has become known as the `candidate selection problem'. In response, numerous graphical and automated selection methods were developed  \citep{Johnston:1992:lr,Manchester:2001:fo,Edwards:2001:tm,Navarro:2003:ab,Keith:2009jo}, designed to filter candidates in bulk. The filtering procedure used ranged in complexity from a simple signal-to-noise ratio (S/N) cut, through to more complex functions \citep{Lee:2013:sk}. In either case, automated approaches enabled large numbers of candidates to be selected at speed in a reproducible way.\newline

Despite these advances the increasing number of candidates produced by contemporary pulsar surveys, tends to necessitate a pass of manual selection upon the candidates selected by software. Many have therefore turned to machine learning methods to build `intelligent' filters \citep{Eatough:2010:uz,Bates:2012:mb,Zhu:2014:ab,Morello:2014:eb}, capable of reducing the dependence on human input. This has achieved some success. However these methods are often developed for a specific pulsar survey search pipeline, making them unsuitable for use with other surveys without modification. As a consequence, new selection mechanisms are often designed and implemented per survey. As more methods continue to emerge, it becomes increasingly unclear which of these best address the candidate selection problem, and under what circumstances. It is also unclear which are best equipped to cope with the trend for increasing candidate numbers, the overwhelming majority of which arise from noise. Existing approaches are not explicitly designed to mitigate noise, rather they are designed to isolate periodic detections. This does not achieve the same effect as explicitly mitigating noise. For example, isolating periodic candidates as potential pulsars, does not necessarily mitigate the impact of periodic noise. Thus it is possible that these techniques will become less effective over time, as noise becomes responsible for an increasing proportion of all candidates detected.\newline

Existing `intelligent' approaches are also ill-equipped to deal with the data processing paradigm shift, soon to be brought about by next-generation radio telescopes. These instruments will produce more data than can be stored, thus survey data processing, including candidate selection, will have to be done on-line in real-time (or close to). In the real-time scenario it is prohibitively expensive to retain all data collected (see Section 4.3.1). It therefore becomes important to identify and prioritise data potentially containing discoveries for storage. Otherwise such data could be discarded and discoveries missed. Thus new techniques are required \citep{Keane:2014:bc} to ensure preparedness for this processing challenge.\newline

In this paper we describe a new candidate selection approach designed for on-line operation, that mitigates the impact of increasing candidate numbers arising from noise. We develop our arguments for producing such a technique in progressive stages. In Section~2 we describe the candidate generation process. We show that improvements in pulsar survey technical specifications have led to increased candidate output, and infer a trend for exponential growth in candidate numbers which we show to be dominated by noise. We also demonstrate why restricting candidate output based on simple S/N cuts, runs the risk of omitting legitimate pulsar signals. The trend in candidate numbers and the ineffectiveness of S/N filters, allows us to identify what we describe as a `crisis' in candidate selection. In Section~3 we review the different candidate selection mechanisms employed during the past fifty years, to look for potential solutions to the issues raised in Section~2. Based on this review, in Section~4, we discuss these methods. We identify how all will be challenged by the transition to on-line processing required by telescopes such as the Square Kilometre Array (SKA), motivating the development of new approaches. In addition we critique the existing features used to describe pulsar candidates, fed as inputs to the machine learning methods employed by many to automate the selection process. In Section 5 we present our own set of eight candidate features, which overcome some of these deficiencies. Derived from statistical considerations and information theory, these features were chosen to maximise the separation between noise and non-noise arising candidates. In Section~6 we describe our new data stream classification algorithm for on-line candidate selection which uses these features. Section~6 also presents classification results that demonstrate the utility of the new approach, and its high level of pulsar recall. Finally in Section~7 we summarise the paper, and comment on how the use of our method has helped to find 20 new pulsars during the LOFAR Tied-Array All-Sky Survey (LOTAAS), though discovery details will be published elsewhere.
\section{Candidate Generation}
\label{sec:CandGen}
Since the adoption of the Fast Fourier Transform (FFT) \citep{Burns:1969:bg,Taylor:1969:jm,Hulse:1974:jh} the general pulsar search procedure has remained relatively unchanged. Signals focused at the receiver of a radio telescope observing at a central frequency $f_{\rm c}$ (MHz), with bandwidth $B$ (MHz), are sampled and recorded at a predetermined rate at intervals of $t_{\rm samp}$ ($\mu s$), chosen to maximise sensitivity to the class of signals being searched for. The data are subsequently split in to $n_{\rm chans}$ frequency channels, each of width $\Delta v$ (kHz). An individual channel contains $s_{\rm tot}$ samples of the signal taken at the interval $t_{\rm samp}$, over an observational period of length $t_{\rm obs}$ seconds, such that $s_{\rm tot} = \frac{t_{\rm obs}}{t_{\rm samp}}$. Each unique observation is therefore representable as an $n_{\rm chans} \times s_{\rm tot}$ matrix $M$.\newline

\begin{figure*}
	\centering
		\includegraphics[scale=0.6]{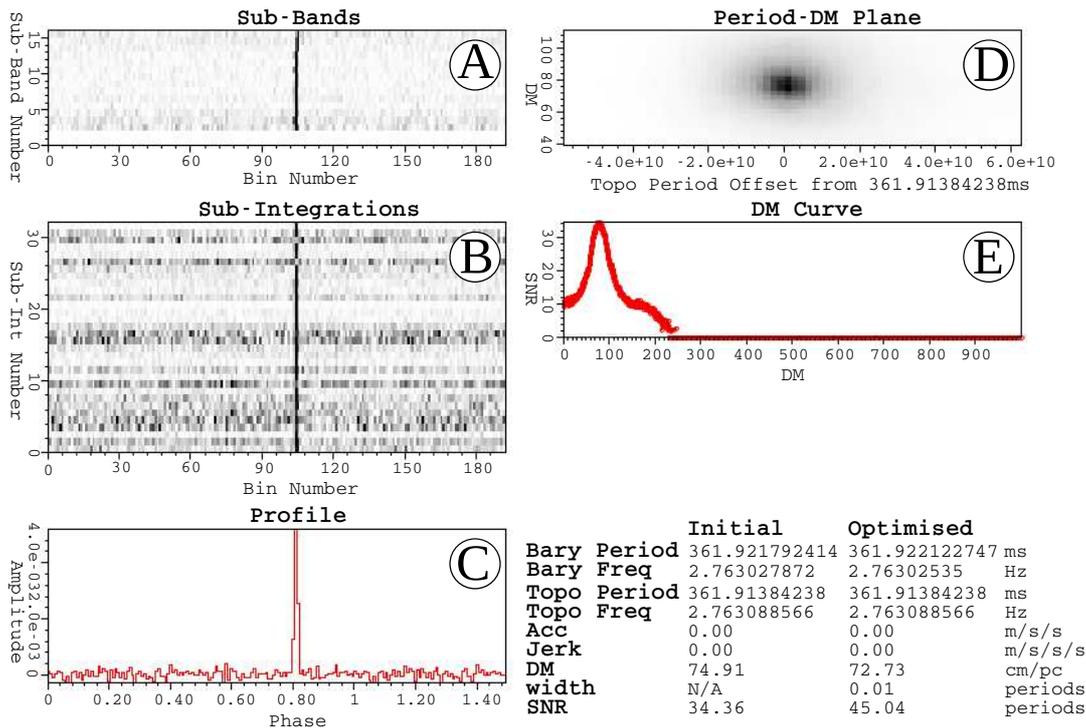}
		\caption[]{An annotated example candidate summarising the detection of PSR J1706-6118. The candidate was obtained during processing of High Time Resolution Universe Survey data by \cite{ThorntonPhD:1}.}
	\label{fig:candidate}
	\vspace{-1.0em}
\end{figure*}
A pulsar search involves a number of procedural steps applied to the data in $M$. The principal steps are
similar for all searches, however the order in which these are undertaken can vary, as too can their precise
implementation. In general, the first step involves radio frequency interference (RFI) excision, via the removal of channels (rows of the matrix) corresponding to known interference frequencies \citep{Keith:2010:bl}. Subsequently `Clipping' \citep{Hogden:2012:vw} may be applied to the data, which aims to reduce the impact of strong interference. This is achieved by setting to zero (or to the local mean) those samples which exhibit intensities higher than some pre-determined threshold in a given column in $M$ (e.g. an intensity $2\sigma$ above the mean). Once these initial steps are complete, processing enters a computationally expensive phase known as de-dispersion.\newline

Dispersion by free electrons in the interstellar medium (ISM) causes a frequency dependent delay in radio emission as it propagates through the ISM. This delay temporally smears legitimate pulsar emission \citep{Lorimer:2005:vm} reducing the S/N of their pulses. The amount of dispersive smearing a signal receives is proportional to a quantity called the Dispersion Measure \citep[DM,][]{Lorimer:2005:vm}. This represents the free electron column density between an observer and a pulsar, integrated along the line of sight. The degree to which a signal is dispersed for an unknown pulsar cannot be known \textit{a priori} \citep[e.g.][]{Keith:2010:bl,LevinPhD:1}, thus several dispersion measure tests or `DM trials' must be conducted to determine this value. This can be used to mitigate the dispersive smearing, thereby increasing the S/N of a signal \citep{Lorimer:2005:vm}. For a single trial, each frequency channel (row in $M$) is shifted by an appropriate delay before each time bin is integrated in frequency. This produces 1 de-dispersed time series for each DM trial value.\newline

Periodic signals in de-dispersed time series data, can be found using a Fourier analysis. This is known as a periodicity search \citep{Lorimer:2005:vm}. The first step after performing the FFT of a periodicity search usually involves filtering the data to remove strong spectral features known as `birdies' \citep{Manchester:2001:fo,Hessels:2007:wt}. These may be caused by periodic or quasi-periodic interference. Summing techniques are subsequently applied, which add the amplitudes of harmonically related frequencies to their corresponding fundamentals. This step is necessary as in the Fourier domain, the power from a narrow pulse is distributed between its fundamental frequency and its harmonics \citep{Lorimer:2005:vm}. Thus for weaker pulsars the fundamental may not rise above the detection threshold, but the harmonic sum generally will. Periodic detections with large Fourier amplitudes post summing (above the noise background or a threshold level), are then considered to be `suspect' periods. \newline

A further process known as sifting \citep[e.g.][]{Stovall:2013:dl} is then applied to the collected suspects, which removes duplicate detections of the same signal at slightly different DMs, along with their related harmonics. A large number of suspects survive the sifting process. Diagnostic plots and summary statistics are computed for each of these remaining suspects forming candidates, which are stored for further analysis. The basic candidate consists of a small collection of characteristic variables. These include the S/N, DM, period, pulse width, and the integrated pulse profile. The latter is an array of continuous variables that describe a longitude-resolved version of the signal that has been averaged in both time and frequency. More detailed candidates also contain data describing how the signal persists throughout the time and frequency domains \citep{Eatough:2010:uz}. This can be seen in plots (A) and (B) in Figure \ref{fig:candidate}. Here persistence in frequency (A) is represented by a two-dimensional matrix showing pulse profiles integrated in time, for a set of averaged frequency channels (i.e. not full frequency resolution). Persistence through time (B), is represented by a two-dimensional matrix showing the pulse profile integrated across similarly averaged frequency channels as a function of time.
\subsection{Modelling Candidate Numbers}
\label{sec:CandidateVolumes}
\begin{table*}
\small
\def\arraystretch{1.3}
\centering
\begin{tabular}{|l|c|c|c|}
\hline
{\bf Survey} & {\bf Year} & {\bf Candidates} & {\bf Per Sq. Degree} \\
\hline\hline
2nd Molonglo Survey\citep{Manchester:1978:rl}				&	1977		&	$2,500$	& $\sim$0.1\\
\hline
Phase II survey \citep{Stokes:1986:sd} 					& 	1983		&	$5,405$	& $\sim$1\\
\hline
Parkes 20 cm survey \citep{Johnston:1992:lr}				& 	1988		&	$\sim150,000$ & $\sim$188\\
\hline
Parkes Southern Pulsar Survey \citep{Manchester:1996:ti}	& 	1991		&	$40,000$	& $\sim$2\\
\hline
Parkes Multibeam Pulsar Survey \citep{Manchester:2001:fo}	& 	1997		&	$8,000,000$	& $\sim$5,161\\
\hline
Swinburne Int. Lat. Survey \citep{Edwards:2001:tm}		&  	1998		&	$>200,000$	& $\sim$168$^{*}$\\
\hline
Arecibo P-Alfa all configurations \citep{Cordes:2006:kh,Lazarus:2012:palfa,PAlfa:2015:pf}						& 	2004		&	$>5,000,000$	& $\sim$16,361$^{*}$\\
\hline
6.5 GHz Multibeam Survey \citep{Bates:2011:sj,BatesPhD:1}	& 	2006		&	$3,500,000$	& $\sim$77,778 \textdagger\\
\hline
GBNCC survey \citep{Stovall:2014:sr}						& 	2009		&	$>1,200,000$	& $\sim$89$^{*}$\\
\hline
Southern HTRU \citep{Keith:2010:bl}				& 	2010		&	$55,434,300$	& $\sim$1,705\\
\hline
Northern HTRU  \citep{Barr:2013:dj,Ng:2012:cn}			& 	2010		&	$>80,000,000$&$\sim$2,890$^{*}$\\
\hline
LOTAAS (Cooper, private communication, 2015 )							& 	2013		&	$39,000,000$	& $\sim$2,000\\
\hline
\end{tabular}
\caption[]{Reported folded candidate numbers. Note $^{*}$ indicates a lower bound on the number of candidates per square degree, calculated from incomplete candidate numbers. $\dagger$  indicates very long integration times, with further details supplied in Tables \ref{tab:specs1} \& \ref{tab:specs2}. }
\label{tab:candidates}
\vspace{-1.0em}
\end{table*}
Candidate numbers are anecdotally understood to be increasing steadily over time. Here we provide historical evidence supporting this view, obtained by reviewing most of the large-scale pulsar surveys conducted since the initial pulsar discovery by \cite{Hewish:1968:jb}. The surveys studied are listed in Tables \ref{tab:specs1} \& \ref{tab:specs2}. This information has also been made available via an interactive on-line resource found at \url{www.jb.man.ac.uk/pulsar/surveys.html}.\newline

Candidate numbers reported in the literature are summarised in Table \ref{tab:candidates}, providing empirical evidence for rising candidate numbers. The rise is understood to be the result of expanding survey technical specifications \citep{Stovall:2013:dl} occurring during the period depicted in Tables \ref{tab:specs1} \& \ref{tab:specs2}. Finer frequency resolution, longer dwell times, and acceleration searches \citep{Eatough:2013:km}, have significantly increased the candidate yield \citep{LyonPhD:1}. However at present there is no accepted method for quantifying the effects of improving survey specifications on candidate numbers. It is therefore difficult to understand precisely how candidate numbers are changing, and what the S/N distribution of candidates should look like in practice. Such knowledge is needed if we are to design candidate selection approaches robust to error, and accurately plan survey storage requirements. Although it is difficult to capture all the steps involved in pulsar data analysis, we describe a model here that can be used as a proxy for estimating candidate numbers, linked to the number of dispersion trials undertaken per observation.
%
%
%
\subsubsection{Approximate Model of Candidate Numbers}
Selection begins in the spectral S/N regime as described in Section \ref{sec:CandGen}. Here each suspect period associated with a spectral S/N, is found through a Fourier analysis of a de-dispersed time series. However, we have incomplete knowledge of the S/N distribution of spectral suspects, which arise from either i) variations in Galactic background noise, ii) RFI, iii) instrument noise, or iv) legitimate phenomena. To overcome this we model only the most significant contributor of candidates, Gaussian distributed background noise. Empirical evidence suggests most candidates originate from background noise. Our analysis of High Time Resolution Universe Survey (HTRU) data  \citep{ThorntonPhD:1} supports this view, held by others \citep{Lee:2013:sk,Morello:2014:eb}. It is also logically consistent, since if most candidates arose from legitimate phenomena discovery would be trivial. Whilst if most arose from RFI, this would be concerning, as telescopes used for surveys are situated in low RFI environments. It thus appears sensible to conclude that candidates are noise dominated.\newline

By modelling candidates arising only from background noise, we can estimate the approximate number of candidates a survey will yield. To achieve this, we assume there is a 1:1 mapping from spectral suspects to folded candidates\footnote{A candidate obtained by folding a de-dispersed time series at a specific suspect period.}. We can then model the folded S/N distribution of noise-originating candidates only, from there onwards. By assuming at least 1 folded candidate is generated per dispersion trial, which also subsequently survives sifting, it is possible to calculate indicative candidate
numbers. As folded candidate S/Ns are empirically well approximated by a Gaussian distribution\footnote{Empirically observed in HTRU survey data.}, we can also estimate the folded S/N distribution using a simple Gaussian model. The number of candidates arising from noise with a folded S/N of  $n\sigma$ (i.e. $1\sigma,...,n\sigma$), is estimated as follows using a Gaussian probability density function,

{\center \begin{gather}\label{eq:f}
f(d,\lambda,\mu,\sigma) = \frac{1}{\sigma \sqrt{2\pi}}{\rm e}^{\rm -\frac{1}{2}(\frac{\lambda-\mu}{\sigma})^{2}} \cdot
d\textrm{,}
\end{gather}
}

where $\lambda$ is the folded S/N, $\mu$ is the mean of the noise distribution, $\sigma$ its standard deviation, and $d$ the total number of dispersion trials. This model considers each dispersion trial to be a single draw from the noise distribution. Thus candidate numbers here are determined by $d$, and not a top $C$ candidate cut, as is often used to limit candidate numbers  \citep[e.g.][]{ThorntonPhD:1}. However since cuts are used in practice to remove weak candidates (arising from noise), we will incorporate them into our model. This is achievable whilst retaining knowledge of the resulting folded S/N distribution for a cut $C\in(0,\infty]$. First we compute the total number of candidates arising from Gaussian distributed noise, with a folded S/N $>$ $n_{\sigma}$ using,

\begin{gather}
\label{eq:k}
k(d,\mu,\sigma,n_{\sigma}) = \int_{\rm n_{\sigma}}^\infty f(d,\lambda,\mu,\sigma) d\lambda\textrm{.}
\end{gather} 

\begin{figure}
	\centering
		\includegraphics[scale=0.62]{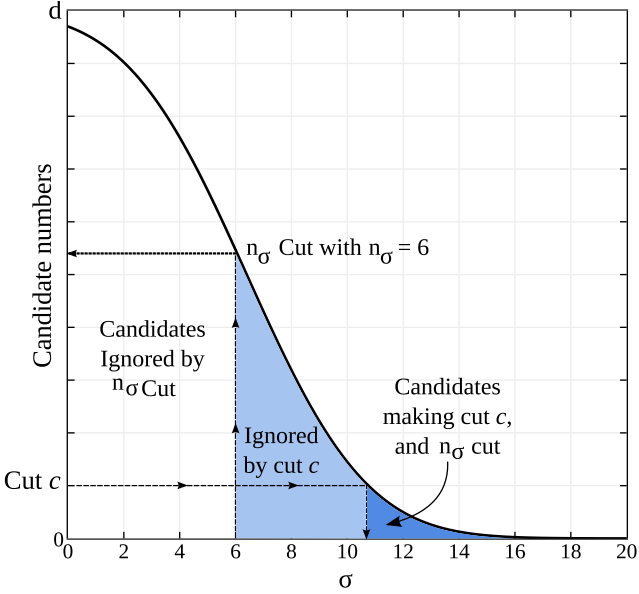}
	\caption[]{Diagram of $1-{\mathrm CDF}$ of Equation 1, showing the relationship between $n\sigma$ and constant cuts. This illustrates their impact on the number of noise candidates making it through to the candidate selection stage.}
	\label{fig:cdf_model}
	\vspace{-1.0em}
\end{figure}

\noindent In practice Gaussian noise possessing a S/N in excess of $30\sigma$ is rare. Thus we can replace the upper limit of $\infty$ with $n_{\rm \sigma}^{\rm max}=30$, beyond which a detection is almost certainly not noise. Here $n_{\rm \sigma}$ is the cut off S/N that is $n$ standard deviations from the mean, and we do not count candidates with S/Ns below this. Equation~\ref{eq:k} is related to the cumulative distribution function (CDF), of the probability distribution in Equation~\ref{eq:f}, where $k = 1 - {\mathrm CDF}$ as shown in Figure \ref{fig:cdf_model}. From this we can compute the number of $n_{\rm \sigma}$ candidates surviving a top $C$ cut, using $h(f,C-k)$. Here $C-k$ gives the number of remaining candidate places in a top $C$ cut, and $h$ is defined by,

\begin{figure}
	\centering
		\includegraphics[scale=0.575]{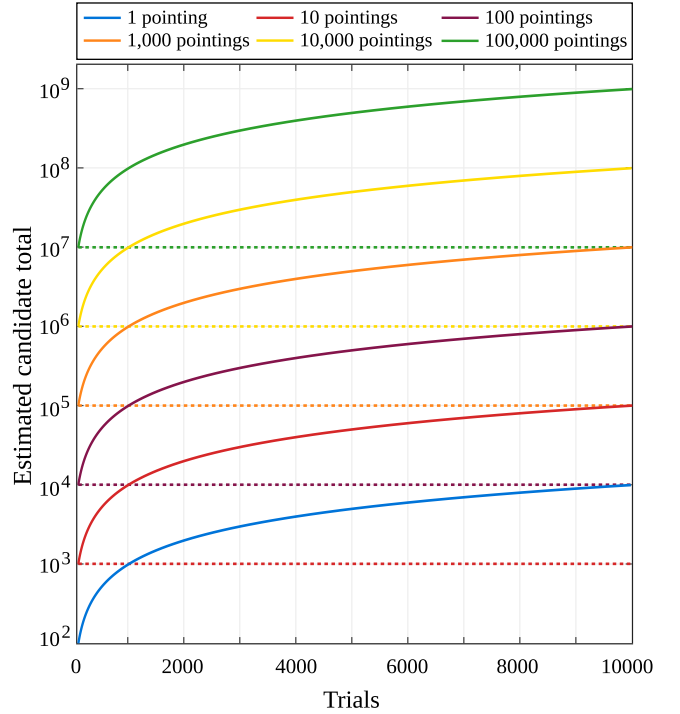}
	\caption[]{Candidate numbers predicted by Equation 4 (using $n_{\rm \sigma}=7$ and $n_{\rm \sigma}^{\rm max}=100$), varied according to the total number of survey pointings for a single beam receiver. Coloured dashed lines indicate the total number of candidates returned when using a conservative $C=100$ cut. The corresponding solid colour lines indicate the total number of candidates returned when the cut is discarded. The solid lines are truncated such that they begin where $C=100$ to avoid overlapping lines complicating the plot.}
	\label{fig:vol_trend}
	\vspace{-1.0em}
\end{figure}

\begin{gather}
\label{eq:h}
h(f,s) = \begin{cases}
      0, & s \leq 0 \\
      f, & f-s \leq 0 \\
      s, & f-s > 0 \textrm{,}
   \end{cases}
\end{gather}

where $0$ is returned if there are no spaces in a top $C$ cut, $f$ is returned if all $n_{\rm \sigma}$ candidates make the cut, and finally $s$ is returned if some $n_{\rm \sigma}$ candidates miss the cut. Now the total number of candidates returned during a survey using a single telescope, with an $n_{\rm beam}$ receiver making $p$ pointings, can be estimated by,

\begin{gather}\label{eq:trend}
p \cdot \big( n_{\rm beams}  \cdot \max (k,C)) \textrm{,}
\end{gather}

where $k$ is given by Equation \ref{eq:k} and $C$ is the numerical cut-off  per beam (e.g. $C=100$). This allows us to identify what we describe as a `crisis' in candidate selection. Since the functions $f$ and $k$ are linearly dependent on $d$, and since we can see empirically from Tables~\ref{tab:specs1} and \ref{tab:specs2} that $d$ is increasing, this means that even if $n_{\rm \sigma}$ is
fixed, the number of noise-originating candidates to be evaluated will increase with $d$. Indeed, Equation \ref{eq:trend} implies the existence of a discernible trend in candidate numbers. Much like the exponential rise in data volumes described by \cite{BatesPhD:1}, this model shows candidate numbers to be increasing exponentially as a function of $d$. This is shown more clearly in Figure \ref{fig:vol_trend}. This illustrates how candidate numbers change as $d$ and the number of survey pointings increase. The plot is colour coded according to total pointings, with dashed lines indicating candidate numbers when $C=100$, and the corresponding solid lines showing candidate numbers when $C$ is discarded. We note here that we have concentrated on utilising $d$ as the number of dispersion trials to show a rise in candidate numbers. This should not be seen simply as relating to the maximum DM being searched. As the sampling time is increased and more channels are used, either to preserve high time resolution at higher dispersion measures, or as the bandwidth increases, or both, then the number of `dispersion' trials increases. Therefore $d$ is also a good proxy for survey sensitivity. Of course the higher the time resolution, the greater $s_{\rm tot}$ increases, longer observations also increase $s_{\rm tot}$ considerably. In both cases this increases the likelihood of detecting more than one candidate per DM trial. For simplicity this is not modelled here, and so what we present can be considered a lower limit.\newline

There are two strategies available for dealing with the implied rise of noisy candidates. The first is to increase
the lower S/N limit $n_{\rm \sigma}$ in Equation \ref{eq:k}. This effectively implements a S/N cut-off, used by many to filter in the spectral domain \citep{Foster:1995:sc,Hessels:2007:wt,Burgay:2013:hk,ThorntonPhD:1}, and the folded domain \citep{Damashek:1978:mt,Manchester:1978:rl,Stokes:1986:sd,Manchester:2001:fo,Burgay:2013:hk}. However in practice this cut-off would become high enough to reject weaker detections of interest (i.e. weaker pulsars, see Section \ref{sec:robust_features}) if it is to reduce candidate numbers. The second option is to impose a smaller constant cut-off $C$ to the candidates collected per observation or beam, also done by many \citep{Edwards:2001:tm,Jacoby:2009:ba,Bates:2012:mb,ThorntonPhD:1} and accounted for in our model. Figure~\ref{fig:cdf_model} shows these two methods to be fundamentally the same. Imposing a fixed limit $C$ on the output of Equation \ref{eq:k}, can only be achieved by increasing the lower value of $n_{\rm \sigma}$ in the integral, since the integrand is fixed by Equation \ref{eq:f}. This corresponds to setting a high S/N cut-off. Using either of these approaches impacts our ability to detect legitimate pulsar signals. This is particularly true of a top $C$ cut, as it would appear that noise alone can fill up a top $C$ cut, without even taking into consideration the influence of RFI, or legitimate phenomena. Taking $d$ to the limit increases the certainty that noise will dominate a candidate cut, and reduces the likelihood of weak legitimate signals making it through to analysis. We now turn our attention to determining how to deal with these issues.
\section{Candidate Selection Methods}
\label{sec:cs}
\subsection{Manual Selection}
During the earliest surveys, manual selection involved the inspection of analogue pen chart records for periodic signals \citep{Large:1968:mi,Manchester:1978:rl}. This process was subsequently replaced by digital data inspection, with the adoption of early computer systems. From then on, manual selection involved the inspection `by eye' of digitally produced diagnostic plots describing each candidate. Those found exhibiting pulsar-like characteristics were recorded for analysis, whilst the remainder were ignored (though retained on disk for possible reanalysis).\newline

During the initial period of digitization, pulsar surveys produced very few candidates with respect to modern searches. The 2nd Molonglo survey conducted during the 1970's, produced only $2,500$ candidates in total \citep{Manchester:1978:rl}. These yielded 224 pulsar detections \citep{Manchester:2005:gv}, a hit rate of almost $9$ per cent\footnote{The hit rate of the recent southern HTRU medium latitude search was much lower, at around $0.01$ per cent \citep{LyonPhD:1}.}. Thus during this period manual selection was entirely practical. Soon after however, increasing candidate numbers began to cause problems. The first mention of this within the literature (to the best of our knowledge) was made by \cite{Clifton:1986:tr} regarding Jodrell Survey B. The number of candidates produced during this survey necessitated extensive manual selection on the basis of pulse profile appearance and S/N. Although such heuristic judgements were not new, their explicit mention with respect to candidate selection indicated that a shift in procedure had occurred. Whereas before it was possible to evaluate most, if not all candidates by eye, here it became necessary to expedite the process using heuristics. Contemporary surveys reacting to similar issues imposed high S/N cut-offs to limit candidate numbers directly. The Arecibo Phase II survey used an $8\sigma$ S/N cut, thus only $\sim$5,405 candidates required manual inspection \citep{Stokes:1986:sd}.\newline

The use of heuristics and S/N cuts proved insufficient to deal with candidate number problems. Additional processing steps such as improved sifting were applied in response, and these became increasingly important during this period. However as these measures apply high up the processing pipeline (close to the final data products), their capacity to reduce candidate numbers was limited. Consequently attempts were made to automatically remove spurious candidates lower down the pipeline, with the aim of preventing them ever reaching human eyes. During the Parkes 20-cm survey, two software tools were devised by \cite{Johnston:1992:lr} to achieve this. Together these encapsulated and optimised the general search procedure discussed in Section 2. The software (`MSPFind' and another unnamed tool) was explicitly designed to reduce the quantity of spurious candidates, while maintaining sensitivity to millisecond pulsars (MSPs). Only candidates with a S/N $>8$ were allowed through the pipeline to manual inspection. It is unclear how many candidates required manual inspection, though the number was less than $150,000$ \citep{Johnston:1992:lr}. During the same period, a similar software tool known as the Caltech Pulsar Package \citep{Deich:1994:wt}, was developed for the Arecibo 430 MHz Intermediate Galactic Latitude Survey \citep{Navarro:2003:ab}. These represent some of the earliest efforts to systematise the search process in a reproducible way.
\subsection{Summary Interfaces}\label{sec:summary_interfaces}
The success achieved via low-level filtering and sifting, continued to be undermined by ever-increasing candidate numbers brought about by technological advances. By the late 1990's, manual selection was therefore becoming increasingly infeasible. This spawned many graphical tools, designed to summarise and filter candidates for speedy and concise evaluation. The first of these, \textsc{runview} \citep{Burgay:2006:jd}, was created to analyse data output by the Parkes Multibeam Survey \citep[PMPS,][]{Manchester:2001:fo}. During the Swinburne Intermediate-latitude survey, \cite{Edwards:2001:tm} devised a similar graphical tool that included distributional information of candidate parameters. A later reprocessing of PMPS data for binary and millisecond pulsars, spawned the development of a more sophisticated graphical tool for candidate viewing called \textsc{reaper}. \textsc{reaper} used a dynamic customizable plot \citep{Faulkner:2004:cl} that enabled heuristic judgements of candidate origin to be made using multiple variables. The use of \textsc{reaper} led to the discovery of 128 unidentified pulsars in PMPS data. This corresponds to $\sim 15.4$ per cent of the known pulsars in PMPS data, given that 833 have now been identified \citep{Lorimer:2015:pe}.\newline

 Following the success of \textsc{reaper}, an updated version of the tool called \textsc{jreaper} was developed by \cite{Keith:2009jo}. It incorporated algorithms which assigned numerical scores to candidates based on their parameters, permitting candidate rankings. By ignoring those candidates achieving low rankings, the amount of visual inspection required was reduced. When applied to data gathered during the PMPS, use of \textsc{jreaper} led to the discovery of a further 28 new pulsars \citep{Keith:2009jo}, corresponding to $\sim 3.4$ per cent of known PMPS pulsars. Thus by 2009, summary interfaces had helped find $\sim 18.7$ per cent of all PMPS pulsars illustrating the usefulness of graphical approaches. More recently, web-based candidate viewing systems incorporating similar scoring mechanisms have appeared \citep{Cordes:2006:kh,Deneva:2009:ac,Deneva:2013:ma}. One such tool, The Pulsar Search Collaboratory \citep{Rosen:2010:sh} \footnote{\url{http://pulsarsearchcollaboratory.com} .}, also incorporates human scoring via the input of high school students. Students taking part in the programme have discovered several new pulsars \citep{Rosen:2013:js}. This includes PSR J1930-1852, a pulsar in a double neutron star system \citep{Swiggum:2015:jk}.
\subsection{Semi-automated Ranking Approaches}
\label{sec:AutomatedApproaches}
Semi-automated selection approaches have recently begun to emerge. Amongst the most popular are those employing ranking mechanisms to prioritise promising candidates for human attention. The most notable of these is the \textsc{PEACE} system developed by \cite{Lee:2013:sk}. \textsc{PEACE} describes each candidate via six numerical features, combined linearly to form a candidate score. Ranked candidates are then analysed via graphical viewing tools by students in the Arecibo Remote Command Centre Programme (ARCC). To date \textsc{PEACE} has been used during the Greenbank Northern Celestial Cap Survey \citep[GBNCC,][]{Stovall:2014:sr} and the Northern High Time Resolution Universe Survey \citep[HTRU North,][]{ Ng:2012:cn,Barr:2013:dj}. Periodic and single-pulse candidates obtained during the A0327 survey \citep{Deneva:2013:ma}, were similarly ranked using an algorithm based on \textsc{PEACE}. Over 50 participants (of varying expertise) from four universities, were then invited to view the A0327 candidates via a web-based interface.
\subsection{Automated `Intelligent' Selection}\label{sec:intelligent_app}
Intelligent selection techniques are gaining widespread adoption. The nature of the intelligence arises from the domain of statistical learning theory, more generally known as machine learning (ML). In particular, from a branch of ML known as \textit{statistical classification}. The aim of classification is to build functions that accurately map a set of input data points, to a set of class labels. For pulsar search this means mapping each candidate to its correct label (pulsar or non-pulsar). This is known as candidate \textit{classification}, a form of supervised learning \citep{Mitchell:1997ua,Duda:2000:hp,Bishop:2006:pr}. If $S=\lbrace X_{\rm 1}, \ldots , X_{\rm n} \rbrace $ represents
the set of all candidate data, then $X_{\rm i}$ is an individual candidate represented by variables known as \textit{features}. Features describe the characteristics of the candidate such that $X_{\rm i} = \lbrace X_{\rm i}^{\rm j},...,X_{\rm i}^{\rm m} \rbrace$, where each feature $X_{\rm i}^{\rm j} \in \mathbb{R}$ for $j=1, \ldots , m$. The label $y$ associated with each candidate, may have multiple possible values such that $y \in Y=\lbrace y_{\rm 1}, \ldots, y_{\rm k} \rbrace$ (e.g. millisecond pulsar, RFI, noise etc). However since the goal here is to separate pulsar and non-pulsar candidates, we consider the binary labels $y\in Y=\lbrace -1,1 \rbrace$, where $y_{\rm 1}=-1$ equates to non-pulsar (synonymous with negative) and $y_{\rm 2}=1$ to pulsar (synonymous with positive).\newline

\begin{figure}
	\centering
		\includegraphics[scale=0.5]{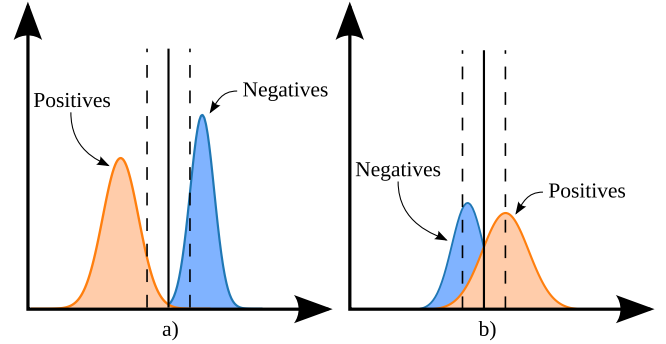}
		\caption[]{Example of the varying separability of features from highly separable in (a), to poorly separable in (b). }
	\label{fig:sep}
	\vspace{-1.0em}
\end{figure}

To build accurate classification systems, it is desirable to utilise features that separate the classes under consideration. This is illustrated in Figure \ref{fig:sep}. An ML function `learns' to separate candidates described using features, from a labelled input vector known as the training set $T$. It contains pairs such that $T=\lbrace (X_{\rm 1},y_{\rm 1}),\ldots,(X_{\rm n},y_{\rm n})\rbrace$. The goal of classification is to induce a mapping function between candidates and labels based on the data in $T$, that minimises generalisation error on test examples \citep{Kohavi:1997:gj}. The derived function can then be used to label new unseen candidates.\newline

The first application of ML approaches to candidate selection was accomplished by \cite{Eatough:2010:uz}. In this work each candidate was reduced to a set of twelve numerical feature values inspired by the scoring system first adopted in \textsc{jreaper}. A predictive model based on a multi-layered perceptron (MLP), a form of artificial neural network \citep{Haykin:1,Bishop:2006:pr}, was then constructed. Using this model, a re-analysis of a sample of PMPS data was completed and a new pulsar discovered \citep{EatoughPhD:1}. Neural network classifiers based on the MLP architecture were also developed to run on data gathered during the HTRU survey. \cite{Bates:2012:mb} modified the earlier approach by describing candidates using 10 further numerical features (22 in total). The same features were used to train neural network classifiers applied to HTRU medium latitude data by \cite{ThorntonPhD:1}. More recently the SPINN system developed by \cite{Morello:2014:eb}, utilised developments from the field of computer science to optimise neural network performance on a set of 6 features. SPINN is currently being applied as part of the Survey for Pulsars and Extragalactic Radio Bursts (SUPERB, Barr 2014; Keane et al. in prep). \newline

Convolutional neural networks \citep[CNN,][]{bengio:2009:jb}, which achieved prominence due to their high accuracy on difficult learning problems such as speech and image recognition, have been adapted for candidate selection. The Pulsar Image-based Classification System (\textsc{pics}) developed by \cite{Zhu:2014:ab}, uses the CNN and other types of machine learning classifier to perform image classification on candidate plots. \textsc{pics} is technically the most sophisticated approach available, and it appears to possess high accuracy. However this comes at the expense of high computational costs. Particularly with respect to runtime complexity. 

\begin{table*}
\small
\centering
\scalebox{0.89}{
\begin{tabular}{|m{5.3cm}|c|c|c|c|c|c|c|c|}
\hline
\textbf{Survey}  & \textbf{Year} & \textbf{$F_{\rm c}$ (MHz)} & \textbf{$B$ (MHz)} & \textbf{$\Delta v$ (kHz)} & \textbf{$n_{\rm chans}$} & \textbf{$t_{\rm samp}(\mu s)$} & \textbf{$t_{\rm obs}(s)$} & \textbf{DM Trials}\\
\hline\hline
1st Molonglo Survey \citep{Large:1968:mi}
&	1968		&	408		&	4		&	2000			&	2		&	5000		&	15			& ?		\\
Search at low Galactic Lat. \citep{Davies:1970:lm}
&	1969		&	408		&	4		&	?			& 	?		& 	50000	&	819			& ?		\\
Arecibo Survey 1 \citep{Hulse:1974:jh}
&	197?		&	430		&	8		&	250			&	32		&	5600		&	198			& 64		\\
Jodrell Survey A \citep{ Davies:1977:sj}
&	1972		&	408		&	4		&	2000			&	2		&	40000	&	660			& ?		\\
2nd Molonglo Survey \citep{Manchester:1978:rl}
&	1977		&	408		&	4		&	800			&	4		&	20000	&	44.7			& ?		\\
Green Bank Northern Hemisphere Survey \citep{Damashek:1978:mt,Damashek:1982:bp}
&	1977		&	400		&	16		&	2000			&	8		&	16700	&	144			& 8		\\
Princeton-NRAO Survey \citep{Dewey:1985:wm}
&	1982-83	&	390		&	16		&	2000			&	8		&	5556		&	138			& 8		\\
Green Bank short-period \citep{Stokes:1985:jh}
&	1983		&	390		&	8		&	250			&	32		&	2000		&	132			& ?		\\
Jodrell Survey B \citep{Clifton:1986:tr}
&	1983-84	&	1400		&	40		&	5000			&	8		&	2000		&	540			& ?		\\
Arecibo survey 2 (a) - Phase II Princeton-NRAO \citep{Stokes:1986:sd}
&	1983		&	390		&	8		&	250			&	32		&	2000		&	132			& ?		\\
Arecibo survey 2 (b) \citep{Stokes:1986:sd}
&	1984-85	&	430		&	0.96		&	60			&	16		&	300		&	39			& ?		\\
Jodrell Survey C \cite{Biggs:1992:ln}
& 1985-87 & {\footnotesize 610/925/928/142}0 & 4/8/32 & 125/500/1000  & 32         &   300     &   79          & 39     \\
Parkes Globular Cluster Survey (20cm) \citep{Manchester:1990:da,Manchester:1990:ra}$^{*}$
&	1989-90	&	1491		&	80/320	&	1000/5000	&	80/64	&	300		&	3000			& 100	\\
Parkes Globular Cluster Survey (50cm) \citep{Manchester:1990:da,Manchester:1990:ra}$^{*}$
&	1988-90	&	640		&	32		&	250			&	128		&	300		&	3000/4500	& 100	\\
Arecibo Survey 3 \citep{Nice:1995:fa}
&	198?		&	430		&	10		&	78.125		&	128		&	516.625	&	67.7			& 256	\\
Parkes 20-cm Survey (I) \citep{Johnston:1992:lr}
&	1988		&	1434		&	800		&	1000			&	80		&	300		&	78.6			& 100	\\
Parkes 20-cm Survey (II) \citep{Johnston:1992:lr}
&	1988		&	1520		&	320		&	5000			&	64		&	1200		&	157.3		& 100	\\
Arecibo 430 MHz Intermediate Galactic Latitude Survey \citep{Navarro:2003:ab}
&	1989-91	&	430		&	10		&	78.125		&	128		&	506.625	&	66.4			& 163	\\
High Galactic Latitude Pulsar Survey of the Arecibo Sky (H1) \citep{Foster:1995:sc}
&	1990		&	430		&	10		&	250			&	128		&	506		&	~40			& 64 	\\
High Galactic Latitude Pulsar Survey of the Arecibo Sky (H2) \citep{Foster:1995:sc}
&	1991		&	430		&	8		&	250			&	32		&	250		&	~40			& 64 	\\
High Galactic Latitude Pulsar Survey of the Arecibo Sky (H3) \citep{Foster:1995:sc}
&	1992		&	430		&	8		&	250			&	32		&	250		&	~40			& 64 	\\
High Galactic Latitude Pulsar Survey of the Arecibo Sky (H4) \citep{Foster:1995:sc}
&	1993		&	430		&	8		&	250			&	32		&	250		&	~40			& 64		\\
High Galactic Latitude Pulsar Survey of the Arecibo Sky (H5) \citep{Foster:1995:sc}
&	1994-95	&	430		&	8		&	250			&	32		&	250		&	~40			& 64		\\
Arecibo Survey 4 Phase I \citep{Nice:1993:jh}
&	1991		&	430		&	10		&	78.125		&	128		&	516.625	&	67.7			& ?		\\
Arecibo Survey 4 Phase II \citep{Camilo:1993:nt}
&	1992		&	429		&	8		&	250			&	64		&	250		&	~40			& 192	\\
Parkes Southern \citep{Manchester:1996:ti}
&	1991-93	&	436		&	32		&	1250			&	256		&	300		&	157.3		& 738	\\
Green Bank fast pulsar survey \citep{Sayer:1997:rw}
&	1994-96	&	370		&	40		&	78.125		&	512		&	256		&	134			& 512	\\
PMPS \citep{Manchester:2001:fo}
&	1997		&	1374		&	288		&	3000			&	96		&	250		&	2100			& 325	\\
Swinburne Int. Lat. survey \citep{Edwards:2001:tm}
&	1998-99	&	1374		&	288		&	3000			&	96		&	125		&	265			& 375	\\
\hline
\end{tabular}
}
\caption{Technical specifications of pulsar surveys conducted between 1968-1999. Here $F_{\rm c}$ (MHz) is the central observing frequency, $B$ (MHz) is the bandwidth, $\Delta v$ (kHz) is the channel width (to 3.d.p), $n_{\rm chans}$ indicates the number of
frequency channels, $t_{\rm samp}(\mu s)$ is the sample frequency (to 3.d.p), and $t_{\rm obs}(s)$ the length of the observation (to 1.d.p). Note * indicates more than one configuration used during the survey. The omission of a survey should be treated as an oversight as opposed to a judgement on its significance.}
\label{tab:specs1}
\vspace{-1.0em}
\end{table*}

\begin{table*}
\small
\centering
\scalebox{0.89}{
\begin{tabular}{|m{6.6cm}|c|c|c|c|c|c|c|c|}
\hline
\textbf{Survey}  & \textbf{Year} & \textbf{$F_{\rm c}$ (MHz)} & \textbf{$B$ (MHz)} & \textbf{$\Delta v$ (kHz)} & \textbf{$n_{\rm chans}$} & \textbf{$t_{\rm samp}(\mu s)$} & \textbf{$t_{\rm obs}(s)$} & \textbf{DM Trials}\\
\hline\hline
Parkes high-lat multibeam \citep{Burgay:2006:jd}
&	2000-03	&	1374		&	288		&	3000			&	96		&	125		&	265			& ?		\\
Survey of the Magellanic Clouds \citep{Manchester:2006:kc}
&	2000-01	&	1374		&	288		&	3000			&	96		&	1000		&	8400			& 228	\\
1.4 GHz Arecibo Survey (DM $<$ 100) \citep{Hessels:2007:wt}
&	2001-02	&	1175		&	100		&	390.625		&	256		&	64		&	7200			& ?		\\
1.4 GHz Arecibo Survey (DM $>$ 100) \citep{Hessels:2007:wt}
&	2001-02	&	1475		&	100		&	195.313		&	512		&	128		&	7200			& ?		\\
Large Area Survey for Radio Pulsars \citep{Jacoby:2009:ba}
&	2001-02	&	1374		&	288		&	3000			&	96		&	125		&	256			& 375	\\
EGRET 56 Pulsar survey \citep{Crawford:2006:fr}
&	2002-03	&	1374		&	288		&	3000			&	96		&	125		&	2100			& 150	\\
EGRET error box survey \citep{Champion:2005:dj}
&	2003		&	327		&	25		&	48.828		&	512		&	125		&	260			& 392	\\
A0327 Pilot \citep{Deneva:2013:ma}
&	2003		&	327		&	25		&	48.828		&	512		&	256		&	60			& 6358	\\
The Perseus Arm Pulsar Survey \citep{Burgay:2013:hk}$^{*}$
&	2004-09	&	1374		&	288		&	3000			&	96		&	125		&	2100			& 183/325\\
The 8gr8 Cygnus Survey \citep{Herrera:2007:gj,Janssen:2009:gh}
&	2004		&	328		&	10		&	19.531		&	512		&	819.2	&	6872			& 488	\\
Parkes deep northern Galactic Plane \citep{Lorimer:2013:mm}
&	2004-05	&	1374		&	288		&	3000			&	96		&	125		&	4200			& 496	\\
P-ALFA Survey (intial) (WAPP) \citep{Cordes:2006:kh,Deneva:2009:ac}
&	2004		&	1420		&	100		&	390.625		&	256		&	64		&	134			& 96		\\
P-ALFA Survey (anticipated) (WAPP) \citep{Cordes:2006:kh,Deneva:2009:ac}$^{*}$
&	2004-10	&	1420		&	300		&	390.625		&	1024		&	64		&	134			& 96/1272\\
6.5 Ghz Multibeam Pulsar Survey \citep{Bates:2011:sj}
&	2006-07	&	6591		&	576		&	3000			&	192		&	125		&	1055			& ~286	\\
Green Bank 350 MHz Drift Scan \citep{Boyles:2013:rl}
&	2007		&	350		&	50		&	24.414		&	2048		&	81.92	&	140			& ?		\\
GBT350 (Spigot) \citep{Deneva:2013:ma}
&	2007		&	350		&	50		&	24.414		&	2048		&	82		&	140			& ?		\\
P-ALFA Survey (MOCK) \citep{Spitler:2014:jm,Deneva:2009:ac,Lazarus:2012:palfa}$^{*}$
&	2009-14	&	1375		&	322.6	&	336.042		&	960		&	65.5		&	120/300		& 5016	\\
GBNCC (GUPPI) \citep{Deneva:2013:ma,Stovall:2014:sr}$^{*}$
&	2009-14	&	350		&	100		&	24.414		&	4096		&	82		&	120			& 17352/26532\\
Southern HTRU (LOW) \citep{Keith:2010:bl}
&	2010-12	&	1352		&	340		&	390.625		&	870		&	64		&	4300			& ?		\\
Southern HTRU (MED) \citep{Keith:2010:bl}
&	2010-12	&	1352		&	340		&	390.625		&	870		&	64		&	540			& 1436	\\
Southern HTRU (HIGH) \citep{Keith:2010:bl}
&	2010-12	&	1352		&	340		&	390.625		&	870		&	64		&	270			& 8000	\\
A0327 (MOCK) \citep{Deneva:2013:ma}
&	2010		&	327		&	57		&	55.664		&	1024		&	125		&	60			& 6358	\\
LPPS \citep{Coenen:2014:tv}
&	2010		&	142		&	6.8		&	12.143		&	560		&	655		&	3420			& 3487	\\
LOTAS \citep{Coenen:2014:tv}$^{*}$
&	2010-11	&	135		&	48		&	12.295		&	3904		&	1300		&	1020			& 16845/18100\\
Northern HTRU (LOW) \citep{Barr:2013:dj,Ng:2012:cn}$^{*}$
&	2010-14	&	1360		&	240		&	585.9		&	410		&	54.61	&	1500			& 406/3240\\
Northern HTRU (MED) \citep{Barr:2013:dj,Ng:2012:cn}$^{*}$
&	2010-14	&	1360		&	240		&	585.9		&	410		&	54.61	&	180			& 406/3240\\
Northern HTRU (HIGH) \citep{Barr:2013:dj,Ng:2012:cn}$^{*}$
&	2010-14	&	1360		&	240		&	585.9		&	410		&	54.61	&	90			& 406/3240\\
SPAN512 \citep{Desvignes:2012:span}
& 	2012		&	1486		&	512		&	500		&	1024		&	64		&	1080			& ?	\\
LOTAAS \citep{lwg:2013:lotaas3,Cooper:2014:lotaas1}
&	2013		&	135		&	95		&	12.207		&	2592		&	491.52	&	3600			& 7000	\\
A0327 (PUPPI) \citep{Deneva:2013:ma}
& 	2014		&	327		&	69		&	24.503		&	2816		&	82		&	60			& 6358	\\
SUPERB (Barr 2014; Keane et al., in prep. )
&	2014		&	1374		&	340		&	332.031		&	1024		&	32		&	540			& 1448		\\
GMRT High Resolution Southern Sky Survey (MID) \citep{Bhattachatyya:2014:bs,Bhattachatyya:2015:bs}
&	2014		&	322		&	32		&	15.625		&	2048		&	60		&	1200			& 6000	\\
GMRT High Resolution Southern Sky Survey (HIGH) \citep{Bhattachatyya:2014:bs,Bhattachatyya:2015:bs}
&	2014		&	322		&	32		&	31.25		&	1024		&	30		&	720			& 6000	\\
FAST* \citep{Smits:2009:el}
&	2016		&	1315		&	400		&	42.105		&	9500		&	100		&	600			& ?		\\
SKA** (Configuration A) \citep{Smits:2009:dc}
&	2020-22	&	1250		&	500		&	50			&	9500		&	64		&	1800			& ?		\\
SKA** (Configuration B) \citep{Smits:2009:dc}
&	2020-22	&	650		&	300		&	50			&	9500		&	64		&	1800			& ?		\\
\hline
\end{tabular}
}
\caption{Technical specifications of pulsar surveys conducted between 2000-present, and projected specifications for instruments under development. X-ray pulsar searches undertaken during this period \citep{Abdo:2009:fl,Ransom:2011:ps} are omitted. Here $F_{\rm c}$ (MHz) is the central observing frequency, $B$ (MHz) is the bandwidth, $\Delta v$ (kHz) is the channel width (to 3.d.p), $n_{\rm chans}$ indicates the number of frequency channels, $t_{\rm samp}(\mu s)$ is the sample frequency (to 3.d.p), and $t_{\rm obs}(s)$ the length of the observation (to 1.d.p). Note * indicates more than one configuration used during the survey. The omission of a survey should be treated as an oversight as opposed to a judgement on its significance.}
\label{tab:specs2}
\vspace{-1.0em}
\end{table*}

\section{Discussion}
\subsection{Critique of Manual Selection}
Manual selection has retained a vital role in pulsar search \citep{Keith:2010:bl}, as demonstrated by its use during recent surveys \citep{Bates:2011:sj,Boyles:2013:rl,Coenen:2014:tv}. The strongest argument in favour of manual selection is its presumed accuracy i.e by \cite{EatoughPhD:1} and \cite{Morello:2014:eb}. However, to the best of our knowledge, no study of the accuracy of expert selection has been conducted. Although intuitively one would expect manual accuracy to be high, studies in other domains indicate otherwise. Most famously studies in medicine and finance \citep{Meehl:1954:pe,Barber:2000:bm} suggest that expert decision-making is flawed due to unconscious biases. Indeed manual selection is already known to be a subjective and error prone process \citep{EatoughPhD:1,Eatough:2010:uz}. In any case, it is infeasible to continue using manual approaches given the rise in candidate numbers predicted in Section \ref{sec:CandidateVolumes}, also anticipated by others \citep{Keane:2014:bc}. Thus irrespective of the true accuracy of manual selection, it must be supplanted to keep pace with increasing data capture rates and candidate numbers.
%
\subsection{Critique of Automated Approaches}
Machine learning approaches are becoming increasingly important for automating decision making processes in finance \citep{Chandola:2009:ba}, medicine \citep{Markou:2003:ss1,Chandola:2009:ba}, safety critical systems \citep{Markou:2003:ss1,Hodge:2004:vaj,Chandola:2009:ba} and astronomy \citep{Borne:2009:wv,Ball:2009:ea,Way:2012:ut}. Given the widespread adoption of ML, the continued application of manual selection raises a fundamental question: why has a transition to completely automated selection not yet occurred? Specific barriers to adoption may be responsible, such as the expertise required to implement and use ML methods effectively. Where this barrier is overcome, approaches emerge that are typically survey and search specific.\newline

A further problem is the limited public availability of pulsar specific code and data. Thus to adopt ML approaches new systems generally need to be built from scratch. Machine learning approaches also have to be `trained' upon data acquired by the same pipeline they will be deployed upon\footnote{The data an algorithm `learns' from must possess the same distribution as the data it will be applied to, otherwise its performance will be poor.}. If training data is not shared, it has to be collected before a survey begins. The cost of doing so may be a further barrier to adoption. Perhaps more simply, existing automated approaches may not yet be accurate enough to be trusted completely. If this is the case, it is unlikely to be caused by the choice of ML system (e.g. neural network, probabilistic classifier, or any other). Those methods described in Section \ref{sec:intelligent_app} employ well studied ML techniques, proven to be effective for a variety of problems. Drops in performance are more likely to be due to deficiencies in i)  the features describing candidates, and ii) the data used to train learning algorithms. In the following section we present evidence suggesting that existing candidate features may well be sub-optimal.
\subsubsection{Sub-optimal Candidate Features}\label{sec:robust_features}
Candidate features can be categorized as being either fundamental to, or as being derived from candidate data. The latter derive new information on the assumption that it will possess some utility, whilst the former do not. For instance the S/N or period of a candidate, can be considered fundamental. A good example of a derived feature is the $\chi^{\rm 2}$ value of a $sine$ curve fit to the pulse profile
 as used by \cite{Bates:2012:mb}. Using curve fittings in this manner expresses an underlying hypothesis. In this case \cite{Bates:2012:mb} suppose a good $\chi^{\rm 2}$ fit to be indicative of sinusoidal RFI. Whilst the reasoning is sound, such a feature represents an untested hypothesis which may or may not hold true.\newline

The majority of existing features are derived \citep[see][]{Eatough:2010:uz,Bates:2012:mb,ThorntonPhD:1,Morello:2014:eb} , and are based upon the heuristics used when selecting candidates manually. As manual selection is imperfect, we cannot rule out the possibility of having designed features, and thereby automated methods, which make the same mistakes as ourselves. Some features in use have been found to introduce unwanted and unexpected biases against particular types of pulsar candidate \citep{Bates:2012:mb,Morello:2014:eb}. Fundamental features are not necessarily better. For example the folded or spectral S/N, is often used as a primitive filter and as a feature for learning. As noise candidates possessing folded S/Ns of $6\sigma$ are common \citep{Nice:1995:fa}, using an S/N cut at this level allows large numbers of likely noise-originating candidates to be rejected. However as noted by \cite{Bates:2012:mb}, such cuts are helpful only if one assumes all low S/N candidates are attributable to noise. In practice the application of cuts has prevented the detection of weaker pulsar signals as warned in Section \ref{sec:CandidateVolumes}. PSR J0812-3910 went unseen in High Latitude survey data \citep{Burgay:2006:jd}, as its spectral S/N was below the survey's threshold for folding. Similarly PSR J0818-3049 went undetected during the same survey, as its folded S/N was below the cut applied prior to manual selection. What's more, there is no agreed upon S/N cut level for any stage in the search pipeline. Domain experience usually plays a role in determining the level, but this is often not specified and difficult to quantify. Levels used include $6\sigma$ \citep{Damashek:1978:mt,ThorntonPhD:1}, $6.3\sigma$ \citep{Manchester:1978:rl}, $7\sigma$ \citep{Foster:1995:sc,Hessels:2007:wt}, $7.5\sigma$ \citep{Manchester:1996:ti}, $8\sigma$ \citep{Stokes:1986:sd,Johnston:1992:lr,Manchester:2001:fo,Edwards:2001:tm,Burgay:2006:jd,Burgay:2013:hk}, $8.5\sigma$ \citep{Nice:1995:fa}, $9\sigma$ \citep{Jacoby:2009:ba,Bates:2011:sj}, and finally $9.5\sigma$ \citep{Jacoby:2009:ba}.\newline\newline
A further problem with many existing features is that they are implementation-dependent. They are described using concepts that can be expressed in various ways mathematically \citep[S/N used by][]{Bates:2011:sj,ThorntonPhD:1,Lee:2013:sk,Morello:2014:eb}, are subject to interpretation without precise definition \citep[pulse width used by][]{Bates:2011:sj,ThorntonPhD:1,Lee:2013:sk,Morello:2014:eb}, or implicitly use external algorithms which go undefined \citep[e.g. curve fitting employed by][]{Bates:2011:sj,ThorntonPhD:1}. It is therefore difficult to build upon the work of others, as features and reported results are not reproducible. Thus direct comparisons between features are rare \citep{Morello:2014:eb} and impractical.
\subsubsection{Feature Evaluation Issues}
The techniques most often used to evaluate features are inadequate for determining how well they separate pulsar and non-pulsar candidates. The most common form of evaluation is undertaken in two steps. The first determines the presence of linear correlations between features and class labels \citep{Bates:2011:sj}, the second compares the performance of different classifiers built using the features \citep{Bates:2011:sj,Lee:2013:sk,Morello:2014:eb} --- the standard `wrapper' method \citep{Kohavi:1997:gj,Guyon:2003:ei}. This  two-step evaluation considers strong linear correlations and accurate classification performance, characteristic of `good' feature sets. However this fails to consider the presence of useful non-linear correlations in the data. Finally using classifier outputs to assess feature performance is known to give misleading results \citep{Brown:2012:ap}, as performance will vary according to the classifier used.\newline\newline
In order to build robust shareable features tolerant to bias, it is necessary to adopt standard procedures that facilitate reproducibility and independent evaluation within the pulsar search community. \cite{Morello:2014:eb} began this process via the sharing of a fully labelled data set, and by providing a clear set of design principles used when creating their features. Here we make similar recommendations, closely followed when designing and evaluating the new feature set described in section \ref{sec:new_features}. It is recommended that features,\newline
\begin{itemize}
\item minimise biases \& selection effects \citep{Morello:2014:eb}
\item be survey-independent for data interoperability
\item be implementation-independent, with concise
\item [] mathematical definitions allowing for reproducibility
\item be evaluated using a statistical framework that enables
\item [] comparison and reproducibility
\item guard against high dimensionality  \citep{Morello:2014:eb}
\item be accompanied by public feature generation code,
\item [] to facilitate co-operation and feature improvement
\item be supplied in a standard data format
\item be evaluated on multiple data sets to ensure robustness.
\end{itemize}
\subsection{Future Processing Challenges}
\begin{figure}
	\centering
		\includegraphics[scale=0.59]{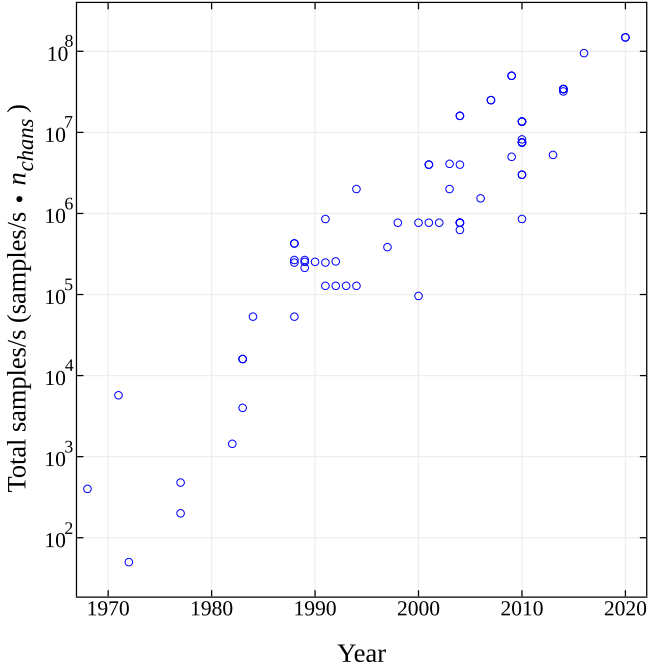}
	\caption{Scatter plot showing the total number of samples per second recorded by all pulsar surveys listed in Tables \ref{tab:specs1} \& \ref{tab:specs2}, as a function of time.}
	\label{fig:SamplesPerSecond}
	\vspace{-1.0em}
\end{figure}
The number of samples per second recorded by pulsar surveys has been increasing steadily over time, as shown in Figure \ref{fig:SamplesPerSecond}. This measure serves as a useful proxy for estimating raw data throughput per second,
\begin{gather}\label{eq:bits_per_sec}
\textrm{bits/s} = \left( \frac{10^{6} }{ t_{\rm samp}} \right) \cdot n_{\rm chans} \cdot n_{\rm pol} \cdot n_{\rm beams} \cdot
n_{\rm bits}\textrm{,}
\end{gather}
where $n_{\rm pol}$ is the number of polarisations, $n_{\rm bits}$ the number of bits used to store an individual sample, and $t_{\rm samp}$ the sampling rate expressed in microseconds. Finer frequency resolution, faster sampling rates and longer observations, increase the data capture rate and thereby the total volume of data generated during a survey. These have been increasing over time as shown in Tables~\ref{tab:specs1} and \ref{tab:specs2}, a trend likely to continue \citep{Smits:2009:dc,Smits:2009:el,Keane:2014:bc}. If it does continue, it will become infeasible to store all raw observational data permanently. It will similarly become impractical to store all candidate data. This is perhaps best illustrated via an example SKA scenario. Suppose for a single observation there are $1,500$ beams and $4,000$ DM trials. If just one candidate is above the S/N selection threshold per DM-acceleration combination, this leads to $4,000$ candidates produced per beam and $6\times 10^{\rm 6}$ per observation. If each candidate is $50$kB in size\footnote{Existing surveys already produce candidates larger than this \citep{Cooper:2014:lotaas1,Barr:2014:Superb,Bhattachatyya:2015:bs}.} then $0.3$TB of candidate data will be generated per observation. For a hypothetical survey lasting 50 days, where there are 120 observations per day, this equates to $3.6 \times 10^{\rm 10}$ individual candidates, and $1.8$PB of candidate data alone (raw data storage requirements are much greater). In the absence of sufficient archiving capacity, here it becomes important to find and prioritise candidates of scientific value for storage. To achieve this, the processing of observational data will have to be done in real-time, from candidate generation to candidate selection. Given the real-time constraint it is impossible to incorporate human decision making into the candidate selection process, thus automated approaches will have to be trusted to accurately determine which data to retain, and which to discard. This will need to be done at high levels of data throughput, with a strict execution time constraint (i.e. before more data arrives). The machine learning methods currently used for candidate filtering as described in Section 3, are not optimised for real-time operation. Rather they are designed for high accuracy, and as such their learning models are not designed to be resource efficient. Their memory and runtime requirements typically grow linearly with the number of candidates observed, whilst quadratic growth or worse is also common. In environments with high data rates, these filters can quickly become processing bottlenecks as their runtime increases. Increasing data rates therefore present two distinct problems for candidate selection: they make it implausible to store all observational data reducing the feasibility of off-line analysis, and restrict our use of candidate selection approaches to those that can operate within strict real-time constraints.
\newline

The shift to on-line processing has already occurred in other domains in response to similar data pressures  \citep{Aad:2008:bc}. Indeed closer to home, some pulsar/fast transient searches are already being undertaken with real-time processing pipelines \citep{Thompson:2011:kw,AitAllal:2012:rw,Barr:2014:Superb,Heerden:2014:vh}. Real-time searches for fast radio bursts \citep[FRBs,][]{Lorimer:2007:mm,Keane:2012:bs,Thornton:2013:tb} are also becoming increasingly common \citep{Law:2014:gc,Petroff:2015:mb,Karastergiou:2015:ac}. These concerns are returned to in Section 6.
\section{New Candidate Features}\label{sec:new_features}
The model introduced in Section \ref{sec:CandidateVolumes} implies that candidate numbers are rising exponentially, and increasingly dominated by noise. We aim to address these problems by finding candidate features that maximise the separation between noise and non-noise candidates, reducing the impact of the largest contributor to high candidate numbers. We also seek to minimise the number of features we use, so as to avoid the problems associated with the `curse of dimensionality' \citep{Hughes:1968:gc}, which reduces classification performance. In total we extracted eight new features for this purpose from two components of the typical pulsar candidate following the recommendations of Section \ref{sec:robust_features}. These features are defined in full in Table \ref{tab:new_features}.\newline

The first four are simple statistics obtained from the integrated pulse profile (folded profile). The remaining four similarly obtained from the DM-SNR curve shown in plot (E) in Figure \ref{fig:candidate}. These features are fundamental to the data, are dissociated with any specific hypothesis, and are few in number. Likewise they possess no intrinsic biases, except perhaps resolution, with respect to the number of profile/DM curve bins used to describe a candidate. The chosen features are also survey/implementation-independent, provided integrated profile and DM-SNR curve data has the same numerical range, and the same ``natural'' DM window\footnote{This is defined as the range of DMs around the DM that yields the highest spectral detection for a candidate. The limits of this range are defined by a change in the DM that corresponds to a time delay across the frequency band equivalent to the initial detection period of a candidate.} for candidates output by different surveys.\newline

 "This is defined as the range of DMs around the DM that gives the highest spectral detection significance for the
 candidate. The limits of this range are defined by the change in dispersion measure that corresponds to a time delay
  across the frequency band equivalent to the candidates intial detection period."

\begin{table}
\small
\centering
\begin{tabular}{|m{1.1cm}|m{2.5cm}|c|}
\hline
{\bf Feature} & {\bf Description} & {\bf Definition} \\
\hline\hline
$Prof_{\mu}$ & Mean of the integrated profile $P$. & $\displaystyle \frac{1}{n}\sum_{\rm i=1}^{\rm n} p_{\rm i}$ \\[12pt]
$Prof_{\sigma}$ & Standard deviation of the integrated profile $P$. & $\displaystyle \sqrt{\frac{\sum_{\rm i=1}^{\rm n}(\rm p_{\rm i}-\bar{P})^{\rm 2}}{n-1}}$ \\[5pt]
$Prof_{k}$ & Excess kurtosis of the integrated profile $P$. & $\displaystyle \frac{\frac{1}{n}(\sum_{\rm i=1}^{n}(p_{\rm i}-\bar{P})^{\rm 4})}{(\frac{1}{n}(\sum_{\rm i=1}^{\rm n}(p_{\rm i}-\bar{P})^{\rm 2}))^{\rm 2}}-3$ \\[12pt]
$Prof_{s}$ & Skewness of the integrated profile $P$. & $\displaystyle \frac{\frac{1}{n}\sum_{\rm i=1}^{\rm n}(p_{\rm i}-\bar{P})^{\rm 3}}{\big(\sqrt{\frac{1}{n}\sum_{\rm i=1}^{\rm n}(p_{\rm i}-\bar{P})^{\rm 2}}\big)^3}$ \\[12pt]
$DM_{\mu}$ & Mean of the DM-SNR curve $D$. & $\displaystyle \frac{1}{n}\sum_{\rm i=1}^{\rm n} d_{\rm i}$ \\[12pt]
$DM_{\sigma}$ & Standard deviation of the DM-SNR curve $D$. & $\displaystyle \sqrt{\frac{\sum_{\rm i=1}^{n}(d_{\rm i}-\bar{D})^{\rm 2}}{n-1}}$ \\
$DM_{k}$ & Excess kurtosis of the DM-SNR curve $D$. & $\displaystyle \frac{\frac{1}{n}(\sum_{\rm i=1}^{n}(d_{\rm i}-\bar{D})^{\rm 4})}{(\frac{1}{n}(\sum_{\rm i=1}^{\rm n}(d_{\rm i}-\bar{D})^{\rm 2}))^{\rm 2}}-3$ \\[12pt]
$DM_{s}$ & Skewness of the DM-SNR curve $D$. & $\displaystyle \frac{\frac{1}{n}\sum_{\rm i=1}^{\rm n}(d_{\rm i}-\bar{D})^{\rm 3}}{\big(\sqrt{\frac{1}{n}\sum_{\rm i=1}^{\rm n}(d_{\rm i}-\bar{D})^{\rm 2}}\big)^3}$ \\[12pt]
\hline
\end{tabular}
\caption[]{The eight features derived from the integrated pulse profile $P=\lbrace p_{\rm 1}, \ldots, p_{\rm n} \rbrace$, and the DM-SNR curve $D=\lbrace d_{\rm 1}, \ldots, d_{\rm n} \rbrace$. For both $P$ and $D$, all $p_{\rm i}$ and $d_{\rm i} \in \mathbb{N} $ for $i=1,...,n$.}
 \label{tab:new_features}
\end{table}
These features were selected by returning to first principles with respect to feature design. By incorporating knowledge of the increasing trend in candidate numbers predicted in Section \ref{sec:CandidateVolumes}, potential features were evaluated according to how well they each separated noise and non-noise candidates. Starting with simple lower-order statistics as possible features (mean, mode, median etc.), the ability of each to reject noise was considered statistically via a three-stage process. Higher order statistics and derived features described by \cite{ThorntonPhD:1} were then added to the pool of possible features, and evaluated similarly. Those achieving the best separation, and the best classification results when used together with machine learning classifiers (see Section \ref{sec:classification_performance}), were then selected for use. Thus these features were chosen with no preconceived notions of their suitability or expressiveness. Rather features were chosen on a statistical basis to avoid introducing bias.
\subsection{Feature Evaluation}
There are three primary considerations when evaluating new features. A feature must i) be useful for discriminating between the various classes of candidate, ii) maximise the separation between them, and iii) perform well in practice when used in conjunction with a classification system. Three separate evaluation procedures have therefore been applied to the features listed in Table \ref{tab:new_features}. The first two forms of evaluation are presented in the section that follows, whilst classification performance is described in Section \ref{sec:classification_performance}, to allow for a comparison between standard classifiers and our stream algorithm described in Section \ref{sec:stream_algo}. As features in themselves are without meaning unless obtained from data, we first describe the data sets used during our analysis, before presenting details of the evaluation.
\subsubsection{Data}
\begin{table}
\def\arraystretch{1.3}
\center
\tabcolsep=0.11cm
    \begin{tabular}{|m{3.8cm}|c|c|c|}
    \hline
    Dataset & Examples & Non-pulsars & Pulsars \\ \hline\hline
    HTRU 1 & 91,192 & 89,995 & 1,196 \\
    HTRU 2 & 17,898 & 16,259 & 1,639 \\
    LOTAAS 1 & 5,053 & 4,987 & 66 \\\hline
    \end{tabular}
\caption[Data]{The pulsar candidate data sets used.}
\label{tab:data}
\end{table}
Three separate datasets were used to test the discriminating capabilities of our features. These are summarised in \mbox{Table~\ref{tab:data}}. The first data set (HTRU 1) was produced by \cite{Morello:2014:eb}. It is the first labelled\footnote{Containing correctly labelled pulsar and non-pulsar candidates.} candidate dataset made publicly available. It consists of $1,196$ pulsar and $89,995$ non-pulsar candidates, in pulsar hunter xml files (.phcx files). These candidates were generated from a re-processing of HTRU Medium Latitude data, using the GPU-based search pipeline \textsc{peasoup} (Barr et al., in prep.). The pipeline searched for pulsar signals with DMs from 0 to 400 cm$^{-3}$pc, and also performed an acceleration search between $-50$ to $+50$ m s$^{-2}$. The HTRU 1 candidate sample possesses varied spin periods, duty cycles, and S/Ns.\newline

\begin{figure*}
	\centering
		\includegraphics[scale=0.69]{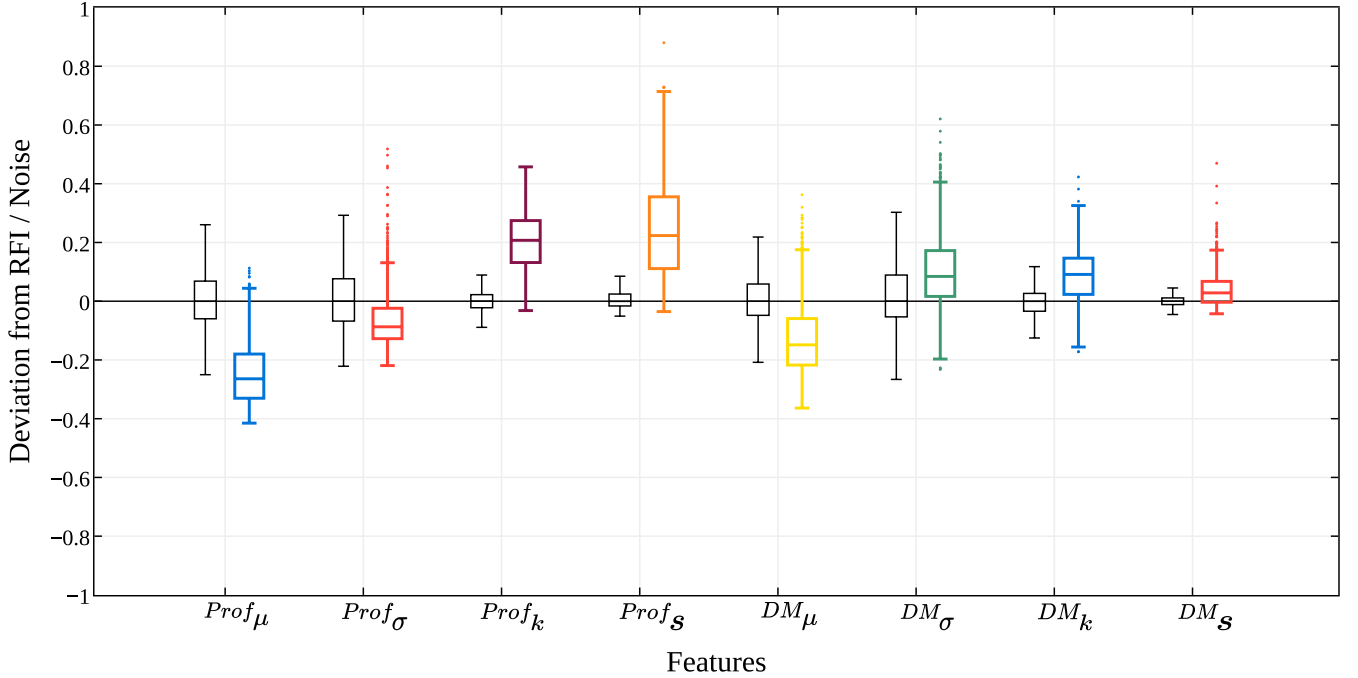}
	\caption{Box plots (median and IQR) showing the linear separability of our new features. Feature data was extracted from $90,000$ labelled pulsar candidates produced by \protect\cite{Morello:2014:eb}, via the \textsc{pulsar feature lab}. There are two box plots per feature. The coloured boxes describe the feature distribution for known pulsars, where corresponding coloured dots represent extreme outliers. Those box plots in black describe the RFI/noise distribution. Note that the data of each feature was scaled to the interval $[0,1]$, before the median of the RFI/noise distribution was subtracted to centre the non-pulsar plots on zero.}
	\label{fig:FeatureSeparation}
	\vspace{-1.0em}
\end{figure*}
In addition two further data sets were used during this work. The first (HTRU 2), is made available for analysis\footnote{\url{https://dx.doi.org/10.6084/m9.figshare.3080389.v1} .}. It comprises $1,639$ pulsar and $16,259$ non-pulsar candidates. These were obtained during an analysis of HTRU Medium Latitude data by \cite{ThorntonPhD:1}, using a search pipeline that searched DMs between 0 to 2000 cm$^{-3}$pc. The pipeline produced over 11 million candidates in total. Of these $1,610$ pulsar and $2,592$ non-pulsar candidates were manually labelled by \cite{Bates:2012:mb} and \cite{ThorntonPhD:1}. These were combined with an additional $13,696$ candidates, sampled uniformly from the same data set according to observational session and month. These additional candidates were manually inspected and assigned their correct labels. Together the two sets of labelled candidates form HTRU 2. It contains 725 of the known 1,108 pulsars in the survey region \citep{LevinPhD:1}, along with re-detections and harmonics. HTRU 2 also contains noise, along with strong and weak forms of RFI. The third and final candidate data set (LOTAAS 1), was obtained during the LOTAAS survey \citep{lwg:2013:lotaas3,Cooper:2014:lotaas1} and is currently private. The data set consists of $66$ pulsar and $4,987$ non-pulsar candidates. Feature data was extracted from these data sets using a new custom written python tool, the \textsc{Pulsar Feature Lab}. This tool is made available for use\footnote{\url{http://dx.doi.org/10.6084/m9.figshare.1536472}}.
\subsubsection{General Separability}\label{sec:corr}
The discriminating capabilities of the new features when applied to HTRU 1, are summarised in Figure \ref{fig:FeatureSeparation} via standard box and whisker plots. For each feature there are two distinct box plots. A coloured box plot representing the feature distribution of known pulsars, and a plain black box plot showing the feature distribution of non-pulsars. As the features have numerical ranges which differ significantly, feature data was scaled to within the range $[0,1]$ prior to plotting. This enables a separability comparison on the same scale. For each individual feature, the median value of the negative distribution was also subtracted. Thus the plots are centred around the non-pulsar median, allowing differences between pulsar and non-pulsar distributions to be seen more clearly.\newline

The visualisation shows there to be a reasonable amount of separation between the pulsar and non-pulsar feature distributions. This is initial evidence for the usefulness of these features\footnote{Similar levels of separability were observed when the same plot was produced for both the HTRU 2 and LOTAAS 1 data sets.} but only on a visual level. Thus we applied a two-tailed students t-test to feature data, in order to determine if the means of the pulsar and non-pulsar distributions were significantly different. A rejection of the null hypothesis (no significant difference) would provide statistical evidence for the separability indicated in the box plots. For all data sets, there was a statistically significant difference between the pulsar and non-pulsar distributions at $\alpha=0.01$. A non-parametric Wilcoxon signed-rank test \citep{Wilcoxon:1945:rs}, was also undertaken with no difference in results. This suggested the features to be worthy of further, more rigorous investigation. The next step involved determining the extent of any linear correlation between the features and the target class variable.
\begin{table}
\centering
\def\arraystretch{1.3}
\begin{tabular}{|c|c|c|c|c|}
\hline
\multirow{2}{*}{Feature} & \multicolumn{3}{c|}{Dataset} & \multirow{2}{*}{Avg. $r_{\rm pb}$} \\\cline{2-4}
 &  HTRU 1     &   HTRU 2    &  LOTAAS 1  &  \\
  \hline\hline
$Prof_{\rm \mu}$ 	& -0.310 & -0.673 & -0.508 &	-0.512 \\\hline
$Prof_{\rm \sigma}$	& -0.084 & -0.364 & -0.337 &	-0.266 \\\hline
$Prof_{\rm k}$		&  0.545 &  0.792 &  0.774 &	 0.719 \\\hline
$Prof_{\rm s}$		&  0.601 &  0.710 &  0.762 &	 0.697 \\\hline
$DM_{\rm \mu}$		& -0.174 &  0.401 &  0.275 &	 0.175 \\\hline
$DM_{\rm \sigma}$	&  0.059 &  0.492 &  0.282 &	 0.287 \\\hline
$DM_{\rm k}$			&  0.178 & -0.391 &  0.426 &	 0.074 \\\hline
$DM_{\rm s}$			&  0.190 & -0.230 & -0.211 &	-0.096 \\\hline
\end{tabular}
\caption[]{The point-biserial correlation coefficient for each feature on the three test data sets.}\label{tab:pbcc}
\end{table}
\subsubsection{Correlation Tests}
The point-biserial correlation coefficient $r_{\rm pb}$ \citep{DasGupta:1960:bp}, measures the linear correlation between variables, when the target variable is dichotomous. It is equivalent to the Pearson product moment \citep{Pearson:1895:pk,Guyon:2003:ei}, though it is better suited to candidate data, as it naturally assumes a discrete target label $y\in Y$ as described previously. The value of $r_{\rm pb}$ for a data sample is given by,
\begin{gather}
r_{pb} = \frac{\bar{x}_{\rm 1}-\bar{x}_{\rm 2}}{\sigma} \cdot \sqrt{ \frac{n_{\rm 2} \cdot n_{\rm 1}}{n \cdot (n-1)} }\textrm{,}
\end{gather}
where $n$ is the total number of samples, $\bar{x}_{\rm 1}$ and $\bar{x}_{\rm 2}$ the mean value of groups one and two respectively, and $\sigma$ the sample standard deviation. Much like Pearson's product moment, the coefficient obtains a value in the range $[-1,1]$. A positive correlation implies that moving from group one to group two, is associated with an increase in the output variable (high values tend to co-occur with group two). A negative correlation implies that moving from group one to group two, is associated with a decrease in the output variable. Table \ref{tab:pbcc} shows the correlation between the eight features and the target class variable, for the three sample data sets. The average (mean) correlation has also been computed. Since $r_{\rm pb}$ is non-additive, this average had to be determined using Fisher's Z transformation \citep{Fisher:1921:ra},
\begin{gather}\label{eq:fishers}
z=\frac{1}{2}\ln\Big(\frac{1+r_{\rm pb}}{1-r_{\rm pb}}\Big) \textrm{.}
\end{gather}
Using Equation \ref{eq:fishers} the corresponding correlations of each feature on the three datasets were transformed into additive $z$ values, summed, and the mean obtained. The mean $z$ value was then transformed back into a meaningful correlation using the inverse of the Fisher-Z,
\begin{gather}
r_{\rm pb}=\frac{e^{\rm 2\cdot z}-1}{e^{\rm 2\cdot z}+1} \textrm{.}
\end{gather}
The data in Table \ref{tab:pbcc} shows there to be three features that on average, exhibit strong correlations ($>|0.5|$). These include the mean, excess kurtosis, and skew of the integrated pulse profile. All features exhibit a  weak correlation on at least one data set, which is stronger on others. The lowest correlation witnessed on HTRU 1, between the standard deviation of the DM-SNR curve and the target variable, performed much better on HTRU 2. This is probably due to differences between the DM ranges of candidates in each dataset (0-400 cm$^{-3}$pc for HTRU 1 and 0-2000 cm$^{-3}$pc for HTRU 2). Irrespective of this no features are completely uncorrelated. Whilst there is variation in the effective linear separability of features across all data sets, it is surprising that such simple measures possess discriminatory ability at all. However, caution must be used when judging features based upon their linear correlations. Those features which possess linear correlations close to zero, may possess useful non-linear correlations which are harder to discern. Thus we turn to the tools of information theory \citep{MacKay:2002:dj,Guyon:2003:ei,Brown:2009:ap} to look for such relationships.
\begin{table*}
\def\arraystretch{1.3}
\begin{tabular}{|c|c|c|c|c|c|c|c|c|}
\hline
\multirow{3}{*}{Feature} & \multicolumn{6}{c|}{Dataset}                                                                 & \multicolumn{2}{c|}{\multirow{2}{*}{Avg.}} \\ \cline{2-7}
& \multicolumn{2}{c|}{HTRU 1} & \multicolumn{2}{c|}{HTRU 2} & \multicolumn{2}{c|}{LOTAAS 1} & \multicolumn{2}{c|}{}\\ \cline{2-9}
& $\displaystyle H(X^{\rm j})$ & $\displaystyle I(X^{\rm j};Y)$ & $\displaystyle H(X^{\rm j})$ & $\displaystyle I
(X^{\rm j};Y)$ & $\displaystyle H(X^{\rm j})$ & $\displaystyle I(X^{\rm j};Y)$ & $\displaystyle H(X^{\rm j})$ & $\displaystyle I(X^{\rm j};Y)$\\ \hline\hline
$Prof_{\rm k}$      & 1.062 & 0.073 & 1.549 & 0.311 & 0.948 & 0.088 & 1.186 & 0.157 \\ \hline
$Prof_{\rm \mu}$    & 1.993 & 0.065 & 2.338 & 0.269 & 1.986 & 0.085 & 2.106 & 0.139 \\ \hline
$Prof_{\rm s}$      & 0.545 & 0.063 & 0.523 & 0.245 & 0.114 & 0.074 & 0.394 & 0.127 \\ \hline
$DM_{\rm k}$        & 1.293 & 0.021 & 2.295 & 0.146 & 1.842 & 0.083 & 1.810 & 0.083 \\ \hline
$Prof_{\rm \sigma}$ & 2.011 & 0.007 & 1.972 & 0.115 & 2.354 & 0.061 & 2.112 & 0.061 \\ \hline
$DM_{\rm \sigma}$   & 2.231 & 0.004 & 2.205 & 0.171 & 0.013 & 0.006 & 1.483 & 0.060 \\ \hline
$DM_{\rm \mu}$      & 1.950 & 0.028 & 0.835 & 0.114 & 0.015 & 0.008 & 0.933 & 0.050 \\ \hline
$DM_{\rm s}$        & 0.138 & 0.013 & 1.320 & 0.041 & 2.243 & 0.045 & 1.233 & 0.033 \\ \hline
\end{tabular}
\caption[]{The entropy $H(X^{\rm j})$, and mutual information $I(X^{\rm j};Y)$ of each feature. Features are ranked
according to their mutual information content with respect to the class label $Y$. Higher mutual information is desireable.}\label{tab:inftheory2}
\end{table*}
\subsubsection{Information Theoretic Analysis}
Information theory uses the standard rules of probability to learn more about features and their interactions. Features which at first appear information-poor, may when combined with one or more other features, impart new and meaningful knowledge \citep{Guyon:2003:ei}. Applying this theory to candidate features enables their comparison, evaluation, and selection within an established framework for the first time.\newline

Information theory describes each feature $X^{\rm j}$ in terms of $entropy$. Entropy is a fundamental unit of information borrowed from Thermodynamics by \citep{Shannon:1949:ce}, that quantifies the uncertainty present in the distribution of $X^{\rm j}$.

The entropy of $X^{\rm j}$ is defined as,
\begin{gather}\label{eq:entropy}
H(X^{\rm j}) = - \sum_{x \in X^{\rm j}} P(x) {\rm \log} _{\rm 2} P(x),
\end{gather}
where $x$ corresponds to each value that $X^{\rm j}$ can take, and $P(x)$ the probability of $x$ occurring. If a given value of $x$ occurs with a high probability, then the entropy of $X^{\rm j}$ is low. Conceptually this can be understood to mean that there is little uncertainty over the likely value of $X^{\rm j}$. Likewise if all possible values of a feature are equally likely, then there is maximum uncertainty and therefore maximum entropy\footnote{Max entropy for a feature with $n$ possible values is given by $\rm log_{2}(n)$.}. Whilst entropy can provide an indication of the uncertainty associated with a feature variable, its
 main usefulness arises when conditioned on the target variable (true class label) $Y$. The conditional entropy of $X^{\rm j}$ given $Y$ is,

\begin{gather}\label{eq:cond_entropy}
H(X^{\rm j}|Y) = - \sum_{y \in Y} p(y) \sum_{x \in X^{\rm j}} P(x|y) {\rm \log} _{\rm 2} P(x|y),
\end{gather}
where $P(x|y)$ is the probability of $x$ given $y$ such that,

\begin{gather}\label{eq:cond_prob}
P(x|y)=\frac{P(x \cap y)}{P(y)}\textrm{.}
\end{gather}
\noindent This quantifies the amount of uncertainty in $X^{\rm j}$ once the value of $Y$ is known. Using Equations 9-11 it is possible to define the mutual information (MI, Brown et al. 2012)\footnote{Also known as information gain, or  a specific case of the Kullback-Leibler divergence \citep{MacKay:2002:dj}.} between the feature $X^{\rm j}$, and the class label $Y$. This can be considered another method of measuring the correlation between a feature and the target variable which detects non-linearities. Mutual information is defined as,
\begin{gather}\label{eq:mi}
I(X^{\rm j};Y)=H(X^{\rm j})-H(X^{\rm j}|Y)\textrm{.}
\end{gather}
The MI expresses the amount of uncertainty in $X^{\rm j}$ removed by knowing $Y$. If $I(X^{\rm j}|Y)=0$ then $X^{\rm j}$ and $Y$ are independent. Whereas if $I(X^{\rm j}|Y)>0$, then knowing $Y$ helps to better understand $X^{\rm j}$. As mutual information is symmetric, knowing $X^{\rm j}$ equivalently helps to better understand $Y$. Thus MI is often described as the amount of information that one variable provides about another \citep{Brown:2012:ap}. It is desirable for features to possess high MI with respect to pulsar/non-pulsar labelling.\newline

The MI metric helps identify relevant features, by enabling them to be ranked according to those that result in the greatest reduction of uncertainty. It is one of the most common filter methods \citep{Kohavi:1997:gj,Guyon:2003:ei,Brown:2012:ap} used for feature selection \citep{Brown:2009:ap}. The entropy and MI of our features are listed in Table \ref{tab:inftheory2}, ranked according to their mean MI content, where higher MI is desirable. To produce this table feature data was discretised, for reasons set out by \citet{Guyon:2003:ei}, enabling use with the information-theoretic \textsc{feast}\footnote{\url{http://www.cs.man.ac.uk/~gbrown/fstoolbox}.} and \textsc{mitoolbox}\footnote{\url{http://www.cs.man.ac.uk/~pococka4/MIToolbox.html}.} toolkits developed by \cite{Brown:2012:ap}. The data was discretised using 10 equal-width bins using the filters within the \textsc{weka} data mining tool\footnote{\url{http://www.cs.waikato.ac.nz/ml/weka}.}. Simple binning was chosen ahead of more advanced Minimum
Description Length (MDL) based discretization procedures \citep{Fayyad:1993:ik}, to simplify feature comparisons.\newline

The four features extracted from the integrated profile contain the largest amounts of MI. These are the most relevant features. The MI content of features extracted from the DM-SNR is much lower. It is tempting therefore to write off these low scoring features since their linear correlation coefficients were also shown to be low in Section \ref{sec:corr}. However whilst mutual information indicates which features are relevant, it is entirely possible for these to contain redundant information \citep{Guyon:2003:ei}. Thus choosing the most relevant features may not produce optimal feature subsets \citep{Kohavi:1997:gj}, since these could contain the same information. The joint mutual information criterion \citep[JMI,][]{Yang:1999:mo} can detect and minimise such redundancy \citep{Guyon:2003:ei,Brown:2012:ap}. Given a set of features the JMI selects those with complementary information, starting with the feature possessing the most mutual information $X^{\rm 1}$. In `forward selection' \citep{Kohavi:1997:gj,Guyon:2003:ei}, a common method of feature selection, a greedy iterative process is used to decide which additional features are most complementary to $X^{\rm 1}$, using the notion of the JMI score,
\begin{gather}\label{eq:jmi}
\textrm{JMI}(X^{\rm j})=\sum_{X^{\rm k}\in F} I(X^{\rm j}X^{\rm k};Y)\textrm{,}
\end{gather}
where $X^{\rm j}X^{\rm k}$ can be understood as a joint probability, and $F$ is the set of features. The iterative process continues until a desired number of features are selected. This produces a feature set that minimises redundancy. Alternatively, if the desired number of features to select equals the total number of those available, features are ranked according to the JMI. Using the JMI in this manner, our features have been ranked such that a lower rank is preferable. Upon applying this criterion poor features are revealed to be useful. This is shown in Table \ref{tab:jmi} which demonstrates that features extracted from the DM-SNR curve impart complementary information, and are therefore ranked higher than profile features which possess greater mutual information. The standard deviation of the DM-SNR curve in particular, is ranked as the 2nd `best' feature on two of the three test datasets. Likewise the excess kurtosis and skewness of the DM-SNR curve, are the second and fourth `best' features for LOTAAS data respectively. In the next section we describe a new data stream classification algorithm, which takes advantage of these features.\newline
\begin{table}
\centering
\def\arraystretch{1.3}
\scalebox{0.97}{
\begin{tabular}{|c|c|c|c|c|}
\hline
\multirow{2}{*}{Feature} & \multicolumn{3}{c|}{Dataset}& \multicolumn{1}{c|}{\multirow{2}{*}{Avg. Rank}} \\ \cline{2-4}
                & HTRU 1 & HTRU 2 & LOTAAS 1& \multicolumn{1}{c|}{}                      \\ \hline\hline
$Prof_{\rm k}$      &	1	 &	  1	   &		1	&  1                                         \\ \hline
$Prof_{\rm \mu}$    &	3	 &	  3	   &		3	&  3                                         \\ \hline
$DM_{\rm \sigma}$   &	2	 &	  2	   &		8	&  4                                         \\ \hline
$Prof_{\rm s}$      &	4	 &	  4	   &		6	& 4.7                                        \\ \hline
$DM_{\rm k}$        &	6	 &	  6	   &		2	& 4.7                                        \\ \hline
$Prof_{\rm \sigma}$ &	7	 &	  5	   &		5	& 5.7                                        \\ \hline
$DM_{\rm \mu}$      &	5	 &	  7	   &		7	& 6.4                                        \\ \hline
$DM_{\rm s}$        &	8	 &	  8	   &		4	& 6.7                                        \\ \hline
\end{tabular}
}
\caption[]{The joint mutual information rank of each feature. Features are ranked according to their average JMI across the three test data sets, where a lower rank is better. }\label{tab:jmi}
\end{table}
\section{Stream Classification}\label{sec:stream_algo}
Data streams are quasi-infinite sequences of information, which are temporally ordered and indeterminable in size \citep{Gaber:2005:ak,Lyon:2013:jk,Lyon:2014:jk}. Data streams are produced by many modern computer systems \citep{Gaber:2005:ak} and are likely to arise from the increasing volumes of data output by modern radio telescopes, especially the SKA. However many of the effective supervised machine learning techniques used for candidate selection do not work with streams \citep{Lyon:2014:jk}. Adapting existing methods for use with streams is challenging, it remains an active goal of data mining research \citep{Xindong:2006:yq,Gaber:2007:vf}. Until that goal is realised, new stream-ready selection approaches are required.
\subsection{Unsuitability of Existing Approaches}
Supervised machine learning methods induce classification models from labelled training sets \citep{Mitchell:1997ua,Bishop:2006:pr}. Provided these are large, representative of rare and majority class examples, and independent \& identically distributed (i.i.d.) to the data being classified \citep{Bishop:2006:pr} good classification performance can be expected to result. However the notion of a training set does not exist within a data stream. There are instead two general processing models used for learning.
\label{batch}
\begin{itemize}
\item {\textbf{Batch processing model:}} at time step $i$ a batch $b$ of $n$ unlabelled instances arrives, and is classified using some model trained on batches $b_{\rm 1}$ to $b_{\rm i-1}$. At time $i+1$ labels arrive for batch $b_{\rm i}$, along with a new batch of unlabelled instances $b_{\rm i+1}$ to be classified.
\item {\textbf{Incremental processing model:}} a single data instance arrives at time step $i$ defined as $X_{\rm i}$, and is classified using some model trained on instances $X_{\rm 1}$ to $X_{\rm i-1}$. At time $i+1$ a label arrives for $X_{\rm i}$, along with a new unlabelled instance $X_{\rm i+1}$ to be classified.
\end{itemize}
In both models learning proceeds continually, as labelled data becomes available. This allows for adaptive learning. Standard supervised classifiers simply cannot be trained in this way. Even if they could, the CPU and memory costs of their training phases make them impractical for streams \citep{Gaber:2012:wi}. This was recognised by \cite{Zhu:2014:ab} with respect to their \textsc{pics} system\footnote{\cite{Zhu:2014:ab} indicated efforts are under way to rectify this.}.\newline

Given these problems how should candidate selection be addressed in streams? One may consider training an existing supervised candidate classifier off-line, which could then be applied to a candidate stream. This is a plausible approach, provided the classifier processes each example before the next one arrives. For this to be viable, the classifier must also be trained with data that is i.i.d. with respect to the data in the stream. However data streams are known to exhibit distributional shifts over varying time periods. For example a changing RFI environment can exhibit shifts over both short (minutes/hours), and/or long (days/weeks/years) time-scales. In either case the shifts cause violations of the i.i.d. assumption, a phenomena known as `concept drift' \citep{Widmer:1996:km,Gaber:2005:ak}. To mitigate the impact of drift, adaptive algorithms able to learn from distributional changes are required, as pre-existing training data no longer characterises the post-drift data distribution \citep{LyonPhD:1}. Such algorithms must be capable of completely reconstructing their internal learning models in an efficient manner per each significant distributional shift. Standard supervised learning models are `static', i.e. they remain unchanged once learned. A static classifier applied to streaming data subject to drifts, will exhibit a significant deterioration in classification performance over time \citep{Aggarwal:2004:ui}. This makes standard supervised learning unsuitable for data streams. In the next section we describe our new `intelligent' data stream classifier, which overcomes these deficiencies. 
\begin{figure}
	\centering
		\includegraphics[scale=0.6]{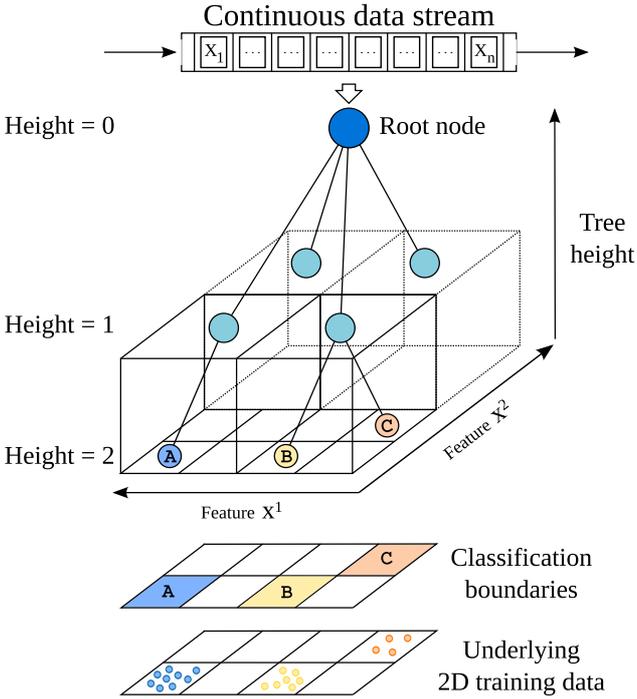}
		\caption[]{An overview of how a streaming decision tree partitions the data space to derive a classification. Each candidate is passed down the tree, and tested at each node it reaches including the root. Each node test outcome determines which branch the candidate continues down, until it reaches a leaf at the bottom of the tree. The tree shown here assigns the class labels A, B and C to examples reaching the leaf nodes.}
	\label{fig:tree}
\end{figure}
\begin{figure*}
	\centering
		\includegraphics[scale=0.5]{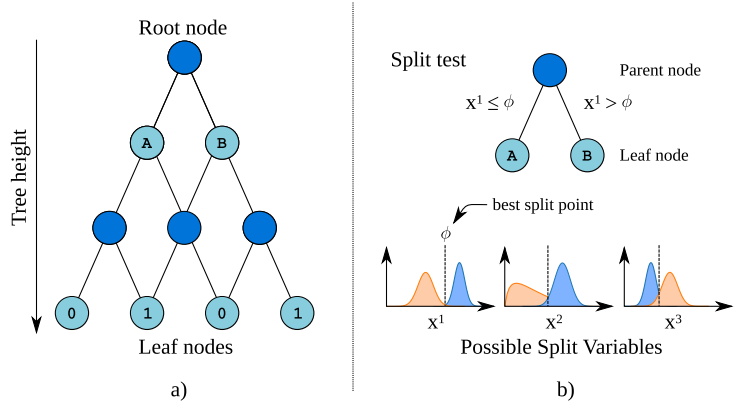}
		\caption[]{An overview of how a decision tree partitions the data space using binary split point `tests' at each node. The best feature variable at each node is first determined, then an optimal numerical split point threshold chosen. Candidates with feature values below the threshold are passed down the left hand branch of the tree, and possibly subjected to further split tests. Similarly for candidates with feature values above the threshold, except these are passed down the right hand branch. Eventually candidates reach the leaf nodes, where they are assigned class labels.}
	\label{fig:split}
\end{figure*}
\subsection{Gaussian-Hellinger Very Fast Decision Tree}
The Gaussian-Hellinger Very Fast Decision Tree (GH-VFDT) is an incremental stream classifier, developed specifically for the candidate selection problem \citep{Lyon:2014:jk}. It is a tree-based algorithm based on the Very Fast Decision tree (VFDT) developed by \cite{Hulten:2001:MTD}. It is designed to maximise classification performance on candidate data streams, which are heavily imbalanced in favour of the non-pulsar class. It is the first candidate selection algorithm designed to mitigate the imbalanced learning problem \citep{Haibo:2009:fb,Lyon:2013:jk,Lyon:2014:jk}, known to reduce classification accuracy when one class of examples (i.e. non-pulsar) dominates the other. The algorithm uses tree learning \citep{Mitchell:1997ua} to achieve this, whereby the data is partitioned using feature split point tests (see Figures \ref{fig:tree} and \ref{fig:split}) that aim to maximise the separation of pulsar and non-pulsar candidates. This involves first choosing the variable that acts as the best class separator, and then finding a numerical threshold `test point' for that variable that maximises class separability.\newline

The tree is `grown' with labelled data to determine optimal splits, using the Hoeffding bound \citep{Hoeffding:1963:ws}. The bound is used to choose statistically with high probability, those split points that would have been selected, if given access to all training data in advance (as in the traditional learning scenario). By calculating the observed mean $\bar{X}^{\rm j}$ of a feature, the bound is able to determine with confidence $1-\delta$ (where $\delta$ is user supplied), that the true mean of the feature is at least $\bar{X}^{\rm j}-\epsilon$ where,
\begin{gather}\label{eq:hoeffding}
\epsilon = \sqrt{\frac{R^{\rm 2} \textrm{ }\ln(1 / \delta)}{2n}}\textrm{,}
\end{gather}
and $R^{\rm 2}$ is the feature range squared. This ensures that the statistically optimal split is always chosen. A split is not made until enough examples in the stream have been seen, i.e. until there is enough evidence to advocate its use. The quality of the splits, and therefore the accuracy of the approach, improve over time. This is because the model of the underlying data distributions improves as more examples are observed. The performance of the algorithm approaches that of a non-streamed classifier as the number of examples observed approaches infinity \citep{Hulten:2001:MTD}. The tree is also able to adapt to change \citep{LyonPhD:1} by updating the data distributions with each observed labelled example. Once there is evidence to suggest an alternative split point is better than one in use, the tree replaces the sub-optimal split. This is achieved by pruning the branch of the tree containing the sub-optimal split, and replacing it with a new branch which begins to `grow' from the new split point.\newline

The key feature of the GH-VFDT, is its use of the skew-insensitive Hellinger distance measure \citep{Hellinger:1909:er,Nikulin:2001:ms} to evaluate split points during learning. This measure makes the classifier robust to the imbalanced learning problem, preventing the classifier from becoming biased towards the abundant non-pulsar class \citep{LyonPhD:1}. By modelling each feature distribution as a Gaussian, the Hellinger distance between the pulsar and non-pulsar distributions can be measured. If $Q$ and $N$ are the pulsar and non-pulsar distributions respectively, the distance for a single feature is given by,
\begin{gather}\label{eq:GaussianHellinger}
d_{H} (Q,N) = \sqrt{1 - \sqrt{ \frac{2 \sigma_{\rm 1} \sigma_{\rm 2}} {\sigma_{\rm 1}^{\rm 2} + \sigma_{\rm 2}^{\rm 2}} } {\rm e}^{ -\frac{1}{4} \frac{ (\mu_{\rm 1}-\mu_{\rm 2})^{\rm 2} } {\sigma_{\rm 1}^{\rm 2} + \sigma_{\rm 2}^{\rm 2}} }} \textrm{,}
\end{gather}
where $Q$ has mean $\mu_{\rm 1}$, variance $\sigma_{\rm 1}^{\rm 2}$ and standard deviation $\sigma_{\rm 1}$, with $N$ defined similarly. The goal of split evaluation is to choose the split which maximises the Hellinger distance, maximising pulsar and non-pulsar separation. This approach requires that only the mean and standard deviation of each feature be known. This significantly reduces the GH-VFDT memory overheads, as knowledge of the entire feature distribution(s) is not
required for learning. Therefore the runtime and memory requirements of the algorithm are sub-linear with respect to the
number of examples processed, and grow in only constant time for each new node added to the tree. This makes the
algorithm suitable for use upon very high throughput data streams such as those described in Section 4.3.\newline

\begin{algorithm}
\small
\caption{Gaussian Hellinger Very Fast Decision Tree}
\begin{algorithmic}[1]
\Require An input stream $	S=\lbrace ..., (X_{\rm i},y_{\rm i}),...\rbrace$, such that each $X_{\rm i}$ is a candidate, $X_{\rm i}^{\rm j}$ its $j$-th feature and $y_{\rm i}$ its class label. The parameter $\delta\in(0,1)$ is the confidence desired, and $\tau\in(0,1)$ a parameter which if set, prevents split point ties.
\Procedure {GH-VFDT}{$S,\delta,\tau$}
\State Let $DT$ be a decision tree with leaf $l_{\rm 1}$
\For{$i \leftarrow 1$ to $|S|$ } \Comment For each stream instance.
	\State $l \leftarrow sort(X_{\rm i},y_{\rm i})$\Comment Sort instance $X_{\rm i}$ to leaf $l$.
	\State $k\leftarrow y_{\rm i}$ \Comment Get class.
	\For{$j \leftarrow 1$ to $|X_{\rm i}^{\rm j}|$ } \Comment For each feature.
		\State update $\mu_{\rm jk}(l,X_{\rm i}^{\rm j})$ \Comment Update observed $\mu$ at leaf.
		\State update $\sigma_{\rm jk}(l,X_{\rm i}^{\rm j})$ \Comment Update observed $\sigma$ at leaf.
	\EndFor
	\State Label $l$ with majority class of instances seen at $l$
	\If{ all $X_{\rm i}$ seen at $l$ don't belong to same class }
	\State $F_{\rm a} \leftarrow null$\Comment Best feature.
	\State $F_{\rm b} \leftarrow null$\Comment 2nd best feature.
		\For{$j \leftarrow 1$ to $|X_{\rm i}^{\rm j}|$ } \Comment For each feature.
		\State $dist\leftarrow d_{H} (X_{\rm i}^{\rm j})$  \Comment From equation \ref{eq:GaussianHellinger}.
		\State $F_{\rm a},F_{\rm b} \leftarrow getBest(dist,X_{\rm i}^{\rm j})$
		\EndFor
		\State $\epsilon = \sqrt{\frac{R^{\rm 2} \textrm{ }ln(1 / \delta)}{2n}}$ \Comment Hoeffding bound.
		\If{ $d_{\rm H} (F_{\rm a})-d_{\rm H} (F_{\rm b}) > \epsilon$ or $\epsilon < \tau$ }
			\State Replace $l$ with new leaf that splits on $F_{\rm a}$
			\For{each branch of split }
				\State Add new leaf $l_{\rm m}$
				\For{$k \leftarrow 1$ to $|S|$ } \Comment For each class.
					\For{$j \leftarrow 1$ to $|X_{\rm i}|$ } \Comment For each $X_{\rm i}^{\rm j}$.
						\State $\mu_{\rm ijk}(l_{\rm m}) \leftarrow 0$
						\State $\sigma_{\rm ijk}(l_{\rm m})  \leftarrow 0$
					\EndFor
				\EndFor
			\EndFor
		\EndIf
	\EndIf
\EndFor
\State \Return $DT$
\EndProcedure
\end{algorithmic}
\end{algorithm}
A complete outline of the GH-VFDT is given in Algorithm 1. On line 7 tree statistics used to compute the Hellinger distance are updated. In particular the running mean and standard deviation maintained at each leaf, for feature $j$, and class $k$ are updated. The call to $getBest(dist,X_{\rm i}^{\rm j})$ returns the best and second best features found at a leaf. This is achieved by choosing those that maximise the Hellinger distance via an iterative process. On line 18 tree split points are first generated and evaluated. Here data is discretized using 10 equal-width bins, and a binary split point chosen.\newline

This approach has already been shown to significantly improve recall rates for pulsar data, above the levels achieved by established stream classifiers. When applied to a data stream containing 10,000 non-pulsar candidates for every legitimate pulsar (HTRU data obtained by \cite{ThorntonPhD:1}), it raised the recall rate from 30 to 86 per cent \citep{Lyon:2014:jk}. This was achieved using candidate data described using the features designed by \cite{Bates:2012:mb} and \cite{ThorntonPhD:1}. A full implementation of the algorithm can be found on-line for public use\footnote{\url{https://github.com/scienceguyrob/GHVFDT}}.
\subsection{Classification Performance}\label{sec:classification_performance}
Existing features and algorithms have been evaluated predominantly in terms of classification accuracy. Such an analysis considers candidate selection as a binary classification problem, whereby candidates arising from pulsars are considered positive (+), and those from non-pulsars negative (-). There are then four possible outcomes for an individual classification decision. These outcomes are summarised in Table \ref{tab:confusionmatrix1} and are evaluated using standard metrics such as those outlined in Table \ref{tab:metrics}. The goal of classification is to minimise the number of false positives, whilst maximising the true positives. Features in this domain are most often chosen according to how well they maximise classifier recall (the fraction of legitimate pulsar candidates correctly classified) and specificity (fraction of non-pulsar candidates correctly classified)\footnote{The approaches in Section \ref{sec:intelligent_app} evaluate in this manner.}. Those classifiers with high recall and specificity exhibit high accuracy, often interpreted to mean that underlying features are good discriminators.\newline
\begin{table}
\def\arraystretch{1.3}
\small
\centering
\begin{tabular}{|m{1.2cm}|m{3cm}|c|}
\hline
{\bf Statistic} & {\bf Description} & {\bf Definition} \\
\hline\hline
Accuracy & Measure of overall classification accuracy. & $\frac{(TP+TN)}{(TP+FP+FN+TN)}$ \\
\hline
False positive rate (FPR) & Fraction of negative instances incorrectly labelled positive. & $\frac{FP}{(FP +TN)}$\\
\hline
G-Mean & Imbalanced data metric describing the ratio between positive and negative accuracy. & $\sqrt{\frac{TP}{TP+FN}\times \frac{TN}{TN+FP}}$\\
\hline
Precision & Fraction of retrieved instances that are positive. & $\frac{TP}{(TP+FP)}$ \\
\hline
Recall & Fraction of positive instances that are retrieved. & $\frac{TP}{(TP + FN)}$\\
\hline
F-Score & Measure of accuracy that considers both precision and recall. & $2\times{ \frac{precision\times{recall}} {precision + recall}} $\\
\hline
Specificity & Fraction of negatives correctly identified as such. & $ \frac{TN}{(FP + TN)}$ \\
\hline
\end{tabular}
\caption[Standard evaluation metrics for classifier performance.]{Standard evaluation metrics for classifier performance. True Positives (TP) are those candidates correctly classified as pulsars. True Negatives (TN) are those correctly classified as \textit{not} pulsars. False Positives (FP) are those incorrectly classified as pulsars, False Negatives (FN) are those incorrectly classified as \textit{not} pulsars. All metrics produce values in the range $[0,1]$. }
 \label{tab:metrics}
 \vspace{-1.0em}
\end{table}

This form of evaluation enables approaches to be tested quickly, with readily interpretable results. However using classifier performance as a proxy to measure feature-separability tests the classification system used as much as the features under investigation \citep{Brown:2012:ap}. The choice of classifier can influence the outcome of the evaluation giving misleading results. Evaluation metrics themselves can also be misleading. Pulsar data sets are imbalanced with respect to the total number of pulsar and non-pulsar candidates within them \citep{Lyon:2013:jk,Lyon:2014:jk}. Thus for data sets consisting of almost entirely non-pulsar examples, high accuracy can often be achieved by classifying all candidates as non-pulsar. In these situations it is an unhelpful metric.\newline

\begin{table}
\def\arraystretch{1.3}
\footnotesize
\centering
\begin{tabular}{ |cc|c|c| }
\hline
\multicolumn{2}{ |c| }{\multirow{2}{*}{}} & \multicolumn{2}{ c| }{\textbf{Predicted}} \\
\cline{3-4}
& & \textit{-} & \textit{+} \\\hline
\multicolumn{1}{ |c }{\multirow{2}{*}{\textbf{Actual}} } &
\multicolumn{1}{ |c| }{\textit{-}} & True Negative (TN) & False Positive (FP) \\ \cline{2-4}
\multicolumn{1}{ |c  }{}                        &
\multicolumn{1}{ |c| }{\textit{+}} & False Negative (FN) & True Positive (TP) \\ \cline{2-4}
\hline
\end{tabular}
\caption[Classifier Metrics.]{Confusion matrix describing the outcomes of binary classification.}
\label{tab:confusionmatrix1}
\vspace{-1.0em}
\end{table}

To overcome these possible sources of inaccuracy when evaluating the GH-VFDT, we make use of the G-mean metric \citep{Haibo:2009:fb}. This describes the ratio between positive and negative accuracy, a measure insensitive to the distribution of pulsar and non-pulsar examples in test data sets. Additionally we employ multiple classifiers in our evaluation which differ greatly in terms of their internal learning models. This allows for a more general view of feature performance in practice to be revealed. This is also useful for evaluating the performance of the GH-VFDT with respect to standard static supervised classifiers, which are at an advantage in such tests. Here we make use of four standard classifiers found in the \textsc{weka} tool. These include the decision tree algorithm C4.5 \citep{Quinlan:1993:c45}, MLP neural network \citep{Haykin:1}, a simple probabilistic classifier Na\"{\i}ve Bayes \citep[NB,][]{Bishop:2006:pr}, and the standard linear soft-margin support vector machine \citep[SVM,][]{Corinna:1995:vp}.
\begin{table*}
\center
\tabcolsep=0.11cm
\scalebox{1.0}{
    \begin{tabular}{|c|c|c|c|c|c|c|c|c|}
    \hline
    Dataset & Algorithm & G-Mean & F-Score & Recall & Precision  & Specificity 	& FPR	& Accuracy \\\hline\hline
\multirow{3}{*}{HTRU 1}
& C4.5 		& 0.962$^*$ 		& 0.839$^*$ 	 	& 0.961 	  	& 0.748  	& 0.962 				& 0.038	& 0.962		\\
& MLP 		& \textbf{0.976} 		& 0.891 	 	& \textbf{0.976} 	  	& 0.820  	& 0.975 			& 0.025$^*$	& 0.975		\\
& NB 		& 0.925 		& 0.837$^*$ 	 	& 0.877 	  	& 0.801  	& 0.975 			& 0.025$^*$	& 0.965		\\
& SVM 	 	& 0.967 		& 0.922 	 	& 0.947 	  	& 0.898  	& 0.988 			& 0.012	& 0.984 \\
& GH-VFDT	& 0.961$^*$ 		& \textbf{0.941} 	 	& 0.928 	  	& \textbf{0.955}  	& \textbf{0.995} 				& \textbf{0.005}	& \textbf{0.988}		\\
\hline\hline
\multirow{3}{*}{HTRU 2}
& C4.5 		& 0.926 		& 0.740 	 	& \textbf{0.904} 	  	& 0.635$^*$  	& 0.949$^*$ 			& 0.051$^*$	& 0.946$^*$		\\
& MLP 		& \textbf{0.931} 		& 0.752 	 	& 0.913 	  	& 0.650$^*$  	& 0.950$^*$ 			& 0.050$^*$	& 0.947$^*$		\\
& NB 		& 0.902 		& 0.692 	 	& 0.863 	  	& 0.579  	& 0.943 			& 0.057	& 0.937		\\
& SVM 	 	& 0.919 		& 0.789 	 	& 0.871 	  	& 0.723  	& 0.969 			& 0.031 & 0.961 \\
& GH-VFDT	& 0.907 		& \textbf{0.862} 	 	& 0.829 	  	& \textbf{0.899}  	& \textbf{0.992} 			& \textbf{0.008}	& \textbf{0.978}		\\
\hline\hline
\multirow{3}{*}{LOTAAS 1}
& C4.5 		& 0.969 		& 0.623 	 	& 0.948 	  	& 0.494  	& 0.991 			& 0.009	& 0.990		\\
& MLP 		& \textbf{0.988} 		& 0.846$^*$ 	 	& \textbf{0.979} 	  	& 0.753  	& 0.998 			& 0.002	& 0.997$^*$		\\
& NB 		& 0.977 		& 0.782 	 	& 0.959 	  	& 0.673  	& 0.996 			& 0.004 & 0.996		\\
& SVM 	 	& 0.949 		& \textbf{0.932} 	 	& 0.901 	  	& \textbf{0.966}  	& \textbf{0.999$^*$} 			& \textbf{0.001$^*$}	& \textbf{0.999}	\\
& GH-VFDT	& 0.888 		& 0.830$^*$ 	 	& 0.789 	  	& 0.875  	& \textbf{0.999$^*$} 			& \textbf{0.001$^*$}	& 0.998$^*$		\\
\hline
    \end{tabular}
    }
    \caption[]{Results obtained on the three test data sets. Bold type indicates the best performance observed. Results with an asterisk indicate no statistically significant difference at the $\alpha=0.01$ level.}
    \label{tab:classifierResults}
    \vspace{-1.0em}
\end{table*}
\subsubsection{GH-VFDT Classification Evaluation Procedure}
Feature data was extracted from the data sets listed in Table \ref{tab:data}, and then independently sampled 500 times. Each sample was split into test and training sets. For HTRU 1 \& 2, sampled training sets consisted of 200 positive and 200 negative examples, with remaining examples making up the test sets. LOTAAS 1 training sets contained 33 positive examples and 200 negative, with remaining examples similarly making up the test sets. Each classifier (five in total) was then trained upon, and made to classify each independent sample, therefore there were $3\times500\times5=7,500$ tests in total. The performance of each algorithm per data set was then averaged to summarise overall performance. To evaluate classifier performance results, one-factor analysis of variance (ANOVA) tests were performed, where the algorithm used was the factor. Tukey's Honestly Significant Difference (HSD) test \citep{Tukey:1949:jv}, was then applied to determine if differences in results were statistically significant at $\alpha=0.01$. The full results are shown in Table~\ref{tab:classifierResults}.\newline

These results indicate that it is possible to achieve high levels of classifier performance using the features described in Section \ref{sec:new_features}. What's more, the classification results are consistent across all three data sets. Recall rates on all three test data sets are high, with $~98$ per cent recall achieved by the MLP on HTRU 1 and LOTAAS 1 data. High levels of accuracy were observed throughout testing and G-mean scores on HTRU 1 were particularly high. The algorithms also exhibited high levels of specificity and generally low false positive rates. The  exception being the $6$ per cent false positive rate achieved by the NB classifier on HTRU 2 data. This outcome is unremarkable for NB, the simplest classifier tested, as the HTRU 2 data set is populated with noise and borderline candidates. Thus we suggest that these represent the first survey independent features developed for the candidate selection problem.\newline

The results also show that the GH-VFDT algorithm consistently outperformed the static classifiers, in terms of both specificity and false positive return rate. This is a highly desirable outcome for a stream classifier, since assigning positive labels too often will return an unmanageable number of candidates. The classifier does not always predict `non-pulsar' to give this result. It is precise, achieving the best precision on two out of the three data sets. G-mean and recall rates were also high for the GH-VFDT, the latter reaching $92.8$ per cent on HTRU 1 data. The recall rates are lower on the remaining two data sets. However it is worth noting that these data sets are considerably smaller than HTRU 1. This is important, since the performance of the GH-VFDT (and of other stream algorithms) improves as more examples are observed. The lower levels of recall on HTRU 2 and LOTAAS 1 are therefore to be expected given the smaller dataset size. In terms of the usefulness of this algorithm for SKA data streams, the GH-VFDT returns consistently less than 1 per cent of candidates as false positives. This greatly reduces the quantity of candidates to be analysed. The GH-VFDT also classifies candidates rapidly. It classified candidates at a rate of $\sim 70,000$ per second using a single 2.2 GHz Quad Core mobile CPU (Intel Core i7-2720QM Processor) when applied to a larger sample of HTRU 2 data consisting of $~11$ million examples. A discussion of the statistics of the pulsars incorrectly classified by the new methods will be discussed in a future paper.
\section{Summary}
This paper has described the pulsar candidate selection process, and contextualised its almost fifty year history. During this time candidate selection procedures have been continually adapting to the demands of increased data capture rates and rising candidate numbers, which has proven to be difficult. We have contributed a new solution to these problems by demonstrating eight new features useful for separating pulsar and non-pulsar candidates, and by developing a candidate classification algorithm designed to meet the data processing challenges of the future. Together these enable a high fraction of legitimate pulsar candidates to be extracted from test data, with recall rates reaching almost $98$ per cent. When applied to data streams, the combination of these features and our algorithm enable over $90$ per cent of legitimate pulsar candidates to be recovered. The corresponding false positive return rate is less than half a percent. Thus together these can be used to significantly reduce the problems associated with high candidate numbers which make pulsar discovery difficult, and go some way towards mitigating the selection problems posed by next generation radio telescopes such as the SKA. The combination of these features and our classification algorithm has already proven useful, aiding in the discovery of 20 new pulsars in data collected during the LOFAR Tied-Array All-Sky Survey \citep{Cooper:2014:lotaas1}. Details of these discoveries will be provided elsewhere, demonstrating the utility of our contributions in practice.\newline

The features described in this paper are amongst the most rigorously tested in this domain. However whilst we advocate their use on statistical grounds, we do not demonstrate their superiority to other features. Future work will consider how these compare to those used previously, and determine if combining them with those already in use is worthwhile. Thus for the time being it is advisable to construct as large a set of features as possible, and use the tools described herein to select feature sets statistically.
\section{Acknowledgements}
This work was supported by grant EP/I028099/1 from the UK Engineering and Physical Sciences Research Council (EPSRC). HTRU 2 data was obtained by the High Time Resolution Universe Collaboration using the Parkes Observatory, funded by the Commonwealth of Australia and managed by the CSIRO. LOFAR data was obtained with the help of the DRAGNET team, supported by ERC Starting Grant 337062 (PI Hessels). We would also like to thank Konstantinos Sechidis for some insightful discussions with respect to information theoretic feature selection, Dan Thornton for initially processing HTRU 2 data, and our reviewer for their helpful feedback.



\bibliographystyle{mnras}

\begin{thebibliography}{99}


\bibitem[\protect\citeauthoryear{ATLAS Collaboration}{2008}]{Aad:2008:bc} ATLAS Collaboration, 2008, JINST, 3

\bibitem[\protect\citeauthoryear{Abdo et al.}{2009}]{Abdo:2009:fl} {Abdo} A.~A. et al., 2009, Science, 325, 5942, p.840

\bibitem[\protect\citeauthoryear{Aggarwal et al.}{2004}]{Aggarwal:2004:ui} {Aggarwal} C. et al., 2004, Proc. of the tenth Int. Conf. on Knowledge discovery and data mining, p.503-508

\bibitem[\protect\citeauthoryear{Ait-Allal et al.}{2012}]{AitAllal:2012:rw} {Ait-Allal} D., {Weber} R., {Dumez-Viou} C., {Cognard} I., {Theureau} G., 2012, C. R. Physique, 13, 1, p.80


\bibitem[\protect\citeauthoryear{Ball \& Brunner.}{2009}]{Ball:2009:ea} {Ball} N., {Brunner} R.~J., 2009,  Int. J. Mod. Phys. D, 19, 7

\bibitem[\protect\citeauthoryear{Barber \& Odean}{2000}]{Barber:2000:bm} {Barber} B.~M., {Odean} T., 2000, Journal of Finance, 55, 2

\bibitem[\protect\citeauthoryear{Barr et al.}{2013}]{Barr:2013:dj} {Barr} E.~D. et al., 2013, MNRAS, 435, 2234

\bibitem[\protect\citeauthoryear{Barr}{2014}]{Barr:2014:Superb} {Barr} E.~D., 2014, presentation at "Extreme-Astrophysics in an Ever-Changing Universe: Time-Domain Astronomy in the 21st Century", Ierapetra, Crete, 16-20 June 2014. http://www3.mpifr-bonn.mpg.de/div/jhs/Program\_files/ EwanBarrCrete2014.pdf (accessed January 6th, 2016)

\bibitem[\protect\citeauthoryear{Bates et al.}{2011a}]{Bates:2011:sj} {Bates} S.~D. et al., 2011a, MNRAS, 411, 1575

\bibitem[\protect\citeauthoryear{Bates et al.}{2011b}]{Bates:2011:fo} {Bates} S.~D. et al., 2011b, MNRAS, 416, 2455

\bibitem[\protect\citeauthoryear{Bates}{2011}]{BatesPhD:1} {Bates} S.~D., 2011, PhD thesis, Univ. Manchester

\bibitem[\protect\citeauthoryear{Bates et al.}{2012}]{Bates:2012:mb} {Bates} S.~D. et al., 2012, MNRAS, 427, 1052

\bibitem[\protect\citeauthoryear{Bengio}{2009}]{bengio:2009:jb} {Bengio} J., 2009, Foundations and Trends in Machine Learning, 2, p.1-127

\bibitem[\protect\citeauthoryear{Borne.}{2009}]{Borne:2009:wv} {Borne} K.~D., 2009, in Next Generation of Data Mining, CRC Press, p.91-114

\bibitem[\protect\citeauthoryear{Bhattachatyya}{2014}]{Bhattachatyya:2014:bs} {Bhattachatyya} B., 2014, presentation at "Transient Key science project meeting 2014", Manchester, UK, 9-10 September 2014. http://www.jb.man.ac.uk/meetings/transients2014/ pdfs/Bhaswati.pdf (accessed January 6th, 2016)

\bibitem[\protect\citeauthoryear{Bhattachatyya}{2015}]{Bhattachatyya:2015:bs} {Bhattachatyya} B. et al., 2015,
astro-ph/1509.07177

\bibitem[\protect\citeauthoryear{Biggs \& Lyne}{1992}]{Biggs:1992:ln} {Biggs} J.~D., {Lyne}, A.~G., 1992, MNRAS, 254, 257-263

\bibitem[\protect\citeauthoryear{Bishop}{2006}]{Bishop:2006:pr} {Bishop} C.~M., 2006, Pattern Recognition and Machine Learning, Springer

\bibitem[\protect\citeauthoryear{Boyles et al.}{2013}]{Boyles:2013:rl} {Boyles} J. et al., 2013, ApJ, 763, 2


\bibitem[\protect\citeauthoryear{Brown}{2009}]{Brown:2009:ap} {Brown} G., 2009, Twelfth Int. Conf. on A.I. \& Statistics, p.49-56

\bibitem[\protect\citeauthoryear{Brown et al.}{2012}]{Brown:2012:ap} {Brown} G., {Pocock} A., {Zhao} Z., {Luj\'{a}n} M., 2012, JMLR, 13

\bibitem[\protect\citeauthoryear{Burgay et al.}{2006}]{Burgay:2006:jd} {Burgay} M. et al.,2006, MNRAS, 368, 283

\bibitem[\protect\citeauthoryear{Burgay et al.}{2013}]{Burgay:2013:hk} {Burgay} M. et al., 2013, MNRAS, 429, 579

\bibitem[\protect\citeauthoryear{Burns \& Clark}{1969}]{Burns:1969:bg} {Burns} W.~R. and {Clark} B.~G., 1969, A\&A, 2, 280



\bibitem[\protect\citeauthoryear{Camilo et al.}{1993}]{Camilo:1993:nt} {Camilo} F., {Nice} D.~J., {Taylor} J.~H., 1993, ApJ, 412, p.L37

\bibitem[\protect\citeauthoryear{Carilli et al.}{2004}]{Carilli:2004fa} {Carilli} C.~L., {Rawlings}, S., 2004, Science with the Square Kilometre Array, New Astronomy Reviews

\bibitem[\protect\citeauthoryear{Champion et al.}{2005}]{Champion:2005:dj} {Champion} D.~J., {McLaughlin} M.~A., {Lorimer} D.~R., 2005, MNRAS, 364, 1011

\bibitem[\protect\citeauthoryear{Chandola et al.}{2009}]{Chandola:2009:ba} {Chandola} V., {Banerjee} A., and {Kumar} V., 2009, ACM Comp. Surv., 41, 3

\bibitem[\protect\citeauthoryear{Clifton \& Lyne}{1986}]{Clifton:1986:tr} {Clifton} T.~R., {Lyne} A.~G., 1986, Nature, 320, 43

\bibitem[\protect\citeauthoryear{Coenen et al.}{2014}]{Coenen:2014:tv} {Coenen} T. et al., 2014, A\&A, 570, A60

\bibitem[\protect\citeauthoryear{Cooper}{2014}]{Cooper:2014:lotaas1} {Cooper} S., 2014, presentation at LOFAR
Science 2014, Amsterdam, The Netherlands, 7-11 April 2014. http://www.astron.nl/lofarscience2014/Documents/ Tuesday/Session\%20III/Cooper.pdf (accessed January 6th, 2016)

\bibitem[\protect\citeauthoryear{Cordes et al.}{2004}]{Cordes:2004gi} {Cordes} J.~M., {Kramer} M., {Lazio} T.~J.~W., {Stappers} B.~W., {Backer} D.~C., {Johnston} S., 2004, New Astronomy Reviews, 4, 11-12, p.1413

\bibitem[\protect\citeauthoryear{Cordes et al.}{2006}]{Cordes:2006:kh} {Cordes} J.~M. et al., 2006, ApJ, 637, 446

\bibitem[\protect\citeauthoryear{Cortes}{1995}]{Corinna:1995:vp} {Cortes} C., {Vapnik}, V., 1995, ML, 20, 3, p.273

\bibitem[\protect\citeauthoryear{Crawford et al.}{2006}]{Crawford:2006:fr} {Crawford} F. et al., 2006, ApJ, 652, 2, 1499


\bibitem[\protect\citeauthoryear{Damashek et al.}{1978}]{Damashek:1978:mt} {Damashek} M., {Taylor} J.~H., {Hulse} R.~A., 1978, ApJ, 225, L31-L33

\bibitem[\protect\citeauthoryear{Damashek et al.}{1982}]{Damashek:1982:bp} {Damashek} M., {Backus} P.~R., {Taylor} J.~H., {Burkhardt} R.~K., 1982, ApJ, 253, L57-L60

\bibitem[\protect\citeauthoryear{Damour et al.}{1998}]{Damour:1998:eg} {Damour} T., {Esposito-Far\`ese}, G., 1998, Phys. Rev. D, 58, 4

\bibitem[\protect\citeauthoryear{Das Gupta}{1960}]{DasGupta:1960:bp} {Das Gupta} S., 1960, Psychometrika, 25, 4

\bibitem[\protect\citeauthoryear{Davies et al.}{1970}]{Davies:1970:lm} {Davies} J.~G., {Large} M.~I., {Pickwick} A.~C., 1970, Nature, 227

\bibitem[\protect\citeauthoryear{Davies et al.}{1977}]{Davies:1977:sj} {Davies} J.~G., {Lyne} A.~G., {Seiradakis}, J.~H., 1977, MNRAS, 179, 635

\bibitem[\protect\citeauthoryear{Deich}{1994}]{Deich:1994:wt} {Deich} W.~T.~S., 1994, PhD thesis, California Institute of Technology.

\bibitem[\protect\citeauthoryear{Deneva et al.}{2009}]{Deneva:2009:ac} {Deneva} J.~S. et al., 2009, ApJ, 703, 2, 2259

\bibitem[\protect\citeauthoryear{Deneva et al.}{2013}]{Deneva:2013:ma} {Deneva} J.~S., {Stovall} K., {McLaughlin} M.~A., {Bates} S.~D., {Freire} P.~C.~C., {Martinez} J.~G., {Jenet} F., {Bagchi} M., 2013, ApJ, 775, 1

\bibitem[\protect\citeauthoryear{Desvignes et al.}{2012}]{Desvignes:2012:span} {Desvignes} G., {Cognard} I., {Champion} D., {Lazarus} P., {Lespagnol} P., {Smith} D.~A., {Theureau}, G., 2012, IAU Symposium 291, astro-ph/1211.3936 

\bibitem[\protect\citeauthoryear{Dewey et al.}{1985}]{Dewey:1985:wm} {Dewey} R.~J., {Taylor} J.~H., {Weisberg} J.~M., {Stokes} G.~H., 1985, ApJ, 294, 1, L25-L29

\bibitem[\protect\citeauthoryear{Duda et al.}{2000}]{Duda:2000:hp} {Duda} R.~O., {Hart} P.~E., {Stork} D.~G., 2000, Pattern Classification, 2nd Edition


\bibitem[\protect\citeauthoryear{Eatough}{2009}]{EatoughPhD:1} {Eatough} R.~P., 2009, PhD thesis, Univ. Manchester.

\bibitem[\protect\citeauthoryear{Eatough et al.}{2010}]{Eatough:2010:uz} {Eatough} R.~P., {Molkenthin} N., {Kramer} M., {Noutsos} A., {Keith} M.~J., {Stappers} B.~W., {Lyne} A.~G., 2010, MNRAS, 407, 2443

\bibitem[\protect\citeauthoryear{Eatough et al.}{2013}]{Eatough:2013:km} {Eatough} R.~P., {Kramer} M., {Lyne} A.~G., {Keith} M.~J., 2013, MNRAS, 431, 292

\bibitem[\protect\citeauthoryear{Edwards et al.}{2001}]{Edwards:2001:tm} {Edwards} R.~T., {Bailes} M., {van Straten} W., {Britton} M.~C., 2001, MNRAS, 326, 358


\bibitem[\protect\citeauthoryear{Faulkner et al.}{2004}]{Faulkner:2004:cl} {Faulkner} A.~J. et al., 2004, MNRAS, 355, 147

\bibitem[\protect\citeauthoryear{Fayyad \& Irani}{1993}]{Fayyad:1993:ik} {Fayyad} U., {Irani} K., 1993, IJCAI, p.1022-1029

\bibitem[\protect\citeauthoryear{Fisher}{1921}]{Fisher:1921:ra} {Fisher} R.~A., 1921,  Metron 1, 3-32

\bibitem[\protect\citeauthoryear{Flach}{2003}]{Flach:2003:pa} {Flach} P.~A., 2003, Proc. 20th Int. Conf. on ML, p.194-201

\bibitem[\protect\citeauthoryear{Foster et al.}{1995}]{Foster:1995:sc} {Foster} R.~S., {Cadwell} B.~J., {Wolszczan} A., {Anderson} S.~B., 1995, ApJ, 454, 826


\bibitem[\protect\citeauthoryear{Gaber et al.}{2005}]{Gaber:2005:ak} {Gaber} M.~M., {Zaslavsky} A., {Krishnaswamy} S., 2005, ACM SIGMOD Record, 34, 2, p.18-26

\bibitem[\protect\citeauthoryear{Gaber et al.}{2007}]{Gaber:2007:vf} {Gaber} M.~M., {Zaslavsky} A., {Krishnaswamy} S., 2007, Advances in Database Systems, p.39-59

\bibitem[\protect\citeauthoryear{Gaber}{2012}]{Gaber:2012:wi} {Gaber} M.~M., 2012, Wiley Interdisciplinary Reviews: Data Mining and Knowledge Discovery, 2, p.79-85

\bibitem[\protect\citeauthoryear{Guyon \& Elisseeff}{2003}]{Guyon:2003:ei} {Guyon} I., {Elisseeff} A., 2003, JMLR, 3, p.1157-1182


\bibitem[\protect\citeauthoryear{van Heerden et al.}{2014}]{Heerden:2014:vh} {van Heerden} E., {Karastergiou} A., {Roberts} S.~J., {Smirnov} O., 2014, General Assembly and Scientific Symposium (URSI GASS)

\bibitem[\protect\citeauthoryear{Haensel et al.}{2007}]{Haensel:2007gw} {Haensel} P., {Potekhin} A.~Y., {Yakovlev} D.~G., 2007, Astrophysics and Space Science Library, 326

\bibitem[\protect\citeauthoryear{He \& Garcia}{2009}]{Haibo:2009:fb} {He} H., {Garcia} E.~A., 2009, IEEE Transactions on Knowledge and Data Engineering, 21, 9, p.1263-1284

\bibitem[\protect\citeauthoryear{Haykin}{1999}]{Haykin:1} {Haykin} S., 1999, Neural Networks A Comprehensive Foundation, Prentice Hall

\bibitem[\protect\citeauthoryear{Hellinger}{1909}]{Hellinger:1909:er} {Hellinger} E., 1909, Journal f\"{u}r die reine und angewandte Mathematik (Crelle's Journal), 136, p.210-271

\bibitem[\protect\citeauthoryear{Hessels et al.}{2007}]{Hessels:2007:wt} {Hessels} J.~W.~T., {Ransom} S.~M., {Stairs} I.~H., {Kaspi} V.~M., {Freire} P.~C.~C., 2007, ApJ, 670, 363

\bibitem[\protect\citeauthoryear{Hewish et al.}{1968}]{Hewish:1968:jb} {Hewish} A., {Bell} S. J., {Pilkington} J.~D.~H., {Scott} P.~F., {Collins} R.~A., 1968, Nature, 217, 5130

\bibitem[\protect\citeauthoryear{Hodge \& Austin}{2004}]{Hodge:2004:vaj} {Hodge} V. J., {Austin} J., 2004, AI Review, 22, 2

\bibitem[\protect\citeauthoryear{Hoeffding}{1963}]{Hoeffding:1963:ws} {Hoeffding} W., 1963, Journal of the American Statistical Association, 58, 301, p.13-30

\bibitem[\protect\citeauthoryear{Hogden et al.}{2012}]{Hogden:2012:vw} {Hogden} J., {Wiel}, S.~V., {Bower}, G.~C., {Michalak}, S., {Siemion}, A., {Werthimer}, D., 2012, astro-ph.IM/1201.1525

\bibitem[\protect\citeauthoryear{Hughes}{1968}]{Hughes:1968:gc} {Hughes} G., 1968, Information Theory, 14, 1, p.55-63

\bibitem[\protect\citeauthoryear{Hulse \& Taylor}{1974}]{Hulse:1974:jh} {Hulse} R.~A. \& {Taylor} J.~H., 1974, ApJ, 191, L59

\bibitem[\protect\citeauthoryear{Hulten et al.}{2001}]{Hulten:2001:MTD} {Hulten} G., {Spence} L., {Domingos} P., 2001, Proc. of seventh ACM SIGKDD.



\bibitem[\protect\citeauthoryear{Jacoby et al.}{2009}]{Jacoby:2009:ba} {Jacoby} B.~A., {Bailes} M., {Ord} S.~M., {Edwards} R.~T., {Kulkarni} S.~R., 2009, ApJ, 699, 2, p.2009

\bibitem[\protect\citeauthoryear{Janssen et al.}{2009}]{Janssen:2009:gh} {Janssen} G.~H., {Stappers} B.~W., {Braun} R., {van Straten} W., {Edwards} R. T., {Rubio-Herrera} E., {van Leeuwen} J., {Weltevrede} P., 2009, A\& A, 498, 1, p.223-231

\bibitem[\protect\citeauthoryear{Johnston et al.}{1992}]{Johnston:1992:lr} {Johnston} S., {Lyne} A.~G., {Manchester} R.~N., {Kniffen} D.~A., {D'Amico} N., {Lim} J., {Ashworth} M., 1992, MNRAS, 255, 401



\bibitem[\protect\citeauthoryear{Karastergiou et al.}{2015}]{Karastergiou:2015:ac} {Karastergiou} A. et al., 2015, astro-ph.IM/1506.03370

\bibitem[\protect\citeauthoryear{Keane et al.}{2012}]{Keane:2012:bs} {Keane} E.~F., {Stappers}, B.~W., {Kramer} M., {Lyne} A.~G., 2012, MNRAS, 425, L71-L75

\bibitem[\protect\citeauthoryear{Keane et al.}{2014}]{Keane:2014:bc} {Keane} E.~F. et al., E., 2014, Proceedings of Science, PoS(AASKA14)040

\bibitem[\protect\citeauthoryear{Keane \& Petroff}{2015}]{Keane:2015:ep} {Keane} E.~F., {Petroff} E., 2015, MNRAS, 447, 2852

\bibitem[\protect\citeauthoryear{Keith et al.}{2009}]{Keith:2009jo} {Keith} M.~J, {Eatough} R.~P., {Lyne} A.~G., {Kramer} M., {Possenti} A., {Camilo} F., {Manchester} R.~N., 2009, MNRAS, 395, 837

\bibitem[\protect\citeauthoryear{Keith et al.}{2010}]{Keith:2010:bl} {Keith} M.~J. et al., 2010, MNRAS, 409, 619

\bibitem[\protect\citeauthoryear{Knispel et al.}{2013}]{Knispel:2013:et} {Knispel} B. et al., 2013, ApJ, 774, 2

\bibitem[\protect\citeauthoryear{Kohavi \& John}{1997}]{Kohavi:1997:gj} {Kohavi} R., {John} G. H., 1997, A.I., 97, 1-2, p.273-324

\bibitem[\protect\citeauthoryear{Kramer et al.}{2004}]{Kramer:2004vf} {Kramer} M., {Backer} D.~C., {Cordes} J.~M., {Lazio} T.~J.~W., {Stappers} B.~W., {Johnston} S., 2004, New Astronomy Reviews, 48, 11-12, p.993-1002



\bibitem[\protect\citeauthoryear{Large et al.}{1968}]{Large:1968:mi} {Large} M.~I., {Vaughan} A.~E., {Wielebinski} R., 1968, Nature, 220, 5169, p.753

\bibitem[\protect\citeauthoryear{Law et al.}{2014}]{Law:2014:gc} {Law} C.~J. et al., 2014, astro-ph/1412.7536

\bibitem[\protect\citeauthoryear{Lazarus}{2012}]{Lazarus:2012:palfa} {Lazarus} P., 2012, The PALFA Survey, IAU Symposium 291, http://www.pulsarastronomy.net/IAUS291/download/ Oral/IAUS291\_LazarusP.pdf (accessed January 6th, 2016)

\bibitem[\protect\citeauthoryear{Lee et al.}{2012}]{lee:2012:kramer} {Lee} K.~J.,{Guillemot} L., {Yue} Y.~L., {Kramer} M., {Champion} D.~J., 2012, MNRAS, 424, 2832

\bibitem[\protect\citeauthoryear{Lee et al.}{2013}]{Lee:2013:sk} {Lee} K.~J. et al., 2013, MNRAS, 433, 688

\bibitem[\protect\citeauthoryear{Levin}{2012}]{LevinPhD:1} {Levin} L., 2012, PhD thesis, Swinburne University

\bibitem[\protect\citeauthoryear{Lofar Working Group}{2013}]{lwg:2013:lotaas3} {{LOFAR} Pulsar Working Group}, 2013, presentation at LOFAR Status Meeting, Dwingeloo, The Netherlands, March 6th, 2013 http://www.lofar.org/wiki/lib/exe/fetch.php?media= public:lsm\_new:2013\_03\_06\_hesself.pdf (accessed January 6th, 2016)

\bibitem[\protect\citeauthoryear{Lorimer \& Kramer}{2006}]{Lorimer:2005:vm} {Lorimer} D., {Kramer}, M., 2006, Cambridge Univ. Press

\bibitem[\protect\citeauthoryear{Lorimer et al.}{2006}]{Lorimer:2006:ha} {Lorimer} D. et al., 2006, MNRAS, 372, 777

\bibitem[\protect\citeauthoryear{Lorimer et al.}{2007}]{Lorimer:2007:mm} {Lorimer} D.~R., {Bailes} M., {McLaughlin} M.~A., {Narkevic} D.~J., {Crawford} F., 2007, Science, 318, p.777-780

\bibitem[\protect\citeauthoryear{Lorimer et al.}{2013}]{Lorimer:2013:mm} {Lorimer} D.~R., {Camilo} F., {McLaughlin} M.~A., 2013, MNRAS, 434, 347

\bibitem[\protect\citeauthoryear{Lorimer et al.}{2015}]{Lorimer:2015:pe} {Lorimer} D.~R. et al., 2015, astro-ph.IM/1501.05516

\bibitem[\protect\citeauthoryear{Lyon et al.}{2013}]{Lyon:2013:jk} {Lyon} R.~J., {Brooke} J.~M., {Knowles} J.~D., {Stappers} B.~W., 2013, SMC, p.1506-1511

\bibitem[\protect\citeauthoryear{Lyon et al.}{2014}]{Lyon:2014:jk} {Lyon} R.~J., {Brooke} J.~M., {Knowles} J.~D., {Stappers} B.~W., 2014, 22nd International Conference on Pattern Recognition, p.1969-1974

\bibitem[\protect\citeauthoryear{Lyon}{2015}]{LyonPhD:1} {Lyon} R.~J., 2015, PhD thesis, Univ. Manchester


\bibitem[\protect\citeauthoryear{MacKay}{2002}]{MacKay:2002:dj} {MacKay} D.~J.~C., 2002, Information Theory, Inference \& Learning Algorithms

\bibitem[\protect\citeauthoryear{Manchester et al.}{1978}]{Manchester:1978:rl} {Manchester} R.~N., {Lyne} A.~G., {Taylor} J.~H., {Durdin} J.~M., {Large} M.~I., {Little} A.~G., 1978, MNRAS, 185, 409

\bibitem[\protect\citeauthoryear{Manchester et al.}{1990a}]{Manchester:1990:da} {Manchester} R.~N., {Lyne} A.~G., {D'Amico} N., {Johnston} S., {Lim} J., {Kniffen} D.~A., 1990a, Nature, 345, 598

\bibitem[\protect\citeauthoryear{Manchester et al.}{1990b}]{Manchester:1990:ra} {Manchester} R.~N., {Lyne} A.~G., {Robinson} C., {D'Amico} N., {Bailes} M., {Lim} J., 1990b, Nature, 352, 219

\bibitem[\protect\citeauthoryear{Manchester et al.}{1996}]{Manchester:1996:ti} {Manchester} R.~N. et al., 1996, MNRAS, 279, 1235

\bibitem[\protect\citeauthoryear{Manchester et al.}{2001}]{Manchester:2001:fo} {Manchester} R.~N. et al., 2001, MNRAS, 328, 17

\bibitem[\protect\citeauthoryear{Manchester et al.}{2005}]{Manchester:2005:gv} {Manchester} R.~N., {Hobbs} G.~B.,{Teoh}  A., {Hobbs} M., 2005, Astron. J., 129, 4

\bibitem[\protect\citeauthoryear{Manchester et al.}{2006}]{Manchester:2006:kc} {Manchester} R.~N., {Fan} G., {Lyne} A.~G., {Kaspi} V.~M., {Crawford} F., 2006, ApJ, 649, 235

\bibitem[\protect\citeauthoryear{Markou \& Singh}{2003}]{Markou:2003:ss1} {Markou} M., {Singh} S., 2003, Signal Processing, 18, 12, p.2499-2521

\bibitem[\protect\citeauthoryear{Meehl}{1954}]{Meehl:1954:pe} {Meehl} P.~E., 1954, Clinical versus statistical prediction: A theoretical analysis and a review of the evidence

\bibitem[\protect\citeauthoryear{Mickaliger et al.}{2012}]{Mickaliger:2012:dl} {Mickaliger} M.~B. et al., 2012, ApJ, 759, 2

\bibitem[\protect\citeauthoryear{Mitchell}{1997}]{Mitchell:1997ua} {Mitchell} T.~M., 1997, Machine Learning, 1st Edition

\bibitem[\protect\citeauthoryear{Morello et al.}{2014}]{Morello:2014:eb} {Morello} V., {Barr} E.~D., {Bailes} M., {Flynn} C.~M., {Keane} E.~F., {van Straten} W., 2014, MNRAS, 443, 1651


\bibitem[\protect\citeauthoryear{Navarro et al.}{2003}]{Navarro:2003:ab} {Navarro} J., {Anderson} S.~B., {Freire} P.~C.~C., 2003, ApJ, 594, 943

\bibitem[\protect\citeauthoryear{Ng}{2012}]{Ng:2012:cn} {Ng} C., 2012, IAU Symposium, S291, 8, p.53-56

\bibitem[\protect\citeauthoryear{Nice et al.}{1993}]{Nice:1993:jh} {Nice} D.~J., {Taylor} J.~H., {Fruchter} A.~S., 1993, ApJ, 402, L49-L52

\bibitem[\protect\citeauthoryear{Nice et al.}{1995}]{Nice:1995:fa} {Nice} D.~J., {Fruchter} A.~S., {Taylor} J.~H., 1995, ApJ, 449

\bibitem[\protect\citeauthoryear{Nikulin}{2001}]{Nikulin:2001:ms} {Nikulin} N.~S. et al., 2001, Encyclopedia of Mathematics



\bibitem[\protect\citeauthoryear{Pearson}{1895}]{Pearson:1895:pk} {Pearson} K., 1895, R. Soc. Lond., 58, p.347-352

\bibitem[\protect\citeauthoryear{Petroff et al.}{2015}]{Petroff:2015:mb} {Petroff} E. et al., 2015, MNRAS, 447, 246

\bibitem[\protect\citeauthoryear{P-Alfa Consortium}{2015}]{PAlfa:2015:pf} {P-Alfa Consortium}, 2015, web resource, ALFA Pulsar Studies,\newline \url{http://www.naic.edu/alfa/pulsar/} (accessed January 6th, 2016)


\bibitem[\protect\citeauthoryear{Quinlan}{2007}]{Quinlan:1993:c45} {Quinlan} J.~R., 1993,
C4.5: programs for machine learning, Morgan Kaufmann


\bibitem[\protect\citeauthoryear{Ransom et al.}{2011}]{Ransom:2011:ps} {Ransom} S.~M. et al., 2011, ApJ Letters, 727, L16

\bibitem[\protect\citeauthoryear{Rosen et al.}{2010}]{Rosen:2010:sh} {Rosen} R. et al., 2010, The Pulsar Search Collaboratory, Astronomy Education Review, 9, 1

\bibitem[\protect\citeauthoryear{Rosen et al.}{2013}]{Rosen:2013:js} {Rosen} R. et al., 2013, ApJ, 768, 85

\bibitem[\protect\citeauthoryear{Rubio-Herrera et al.}{2007}]{Herrera:2007:gj} {Rubio-Herrera} E., {Braun} R.,{Janssen} G.,{van Leeuwen} J., {Stappers} B.~W., 2007, astro-ph/0701183


\bibitem[\protect\citeauthoryear{Sayer et al.}{1997}]{Sayer:1997:rw} {Sayer} R.~W., {Nice} D.~J., {Taylor} J.~H., 1997, ApJ, 474

\bibitem[\protect\citeauthoryear{Shannon \& Weaver}{1949}]{Shannon:1949:ce} {Shannon} C.~E., {Weaver} W., 1949, The mathematical theory of communication, Univ. of Illinois Press

\bibitem[\protect\citeauthoryear{Smits et al.}{2009a}]{Smits:2009:dc} {Smits} R., {Kramer} M., {Stappers} B., {Lorimer} D.~R., {Cordes} J., {Faulkner} A., 2009a, A\&A, 493, 3, p.1161

\bibitem[\protect\citeauthoryear{Smits et al.}{2009b}]{Smits:2009:el} {Smits} R., {Lorimer} D.~R., {Kramer} M., {Manchester} R., {Stappers} B., {Jin} C.~J., {Nan} R.~D., {Li} D., 2009b, A\&A, 505, 2, p.919-926

\bibitem[\protect\citeauthoryear{Spitler et al.}{2014}]{Spitler:2014:jm} {Spitler} L.~G. et al., 2014, ApJ, 790, 2

\bibitem[\protect\citeauthoryear{Stokes et al.}{1985}]{Stokes:1985:jh} {Stokes} G.~H., {Taylor} J.~H., {Weisberg} J.~M., {Dewey} R.~J., 1985, Nature, 317, p.787-788

\bibitem[\protect\citeauthoryear{Stokes et al.}{1986}]{Stokes:1986:sd} {Stokes} G.~H., {Segelstein} D.~J. {Taylor} J.~H., {Dewey} R.~J., 1986, ApJ, 311, p.694-700

\bibitem[\protect\citeauthoryear{Stovall et al.}{2013}]{Stovall:2013:dl} {Stovall} K., {Lorimer} D.~R., {Lynch} R.~S., 2013, Class. Quantum Grav., 30, 22

\bibitem[\protect\citeauthoryear{Stovall et al.}{2014}]{Stovall:2014:sr} {Stovall} K. et al., 2014, ApJ, 791

\bibitem[\protect\citeauthoryear{Swiggum et al.}{2015}]{Swiggum:2015:jk} {Swiggum} J.~K. et al., 2015, ApJ, 805, 156


\bibitem[\protect\citeauthoryear{Taylor, Dura \& Huguenin}{1969}]{Taylor:1969:jm} {Taylor}, J.~H. and {Jura}, M. and  {Huguenin}, G.~R., 1969,  Nature, 223, 797

\bibitem[\protect\citeauthoryear{Thompson et al.}{2011}]{Thompson:2011:kw} {Thompson} D.~R., {Majid} W.~A., {Wagstaff} K., {Reed} C., 2011,  NASA Conference on Intelligent Data Understanding

\bibitem[\protect\citeauthoryear{Thornton et al.}{2013}]{Thornton:2013:tb} {Thornton} D. et al., 2013, Science, 341, p.53-56

\bibitem[\protect\citeauthoryear{Thornton}{2013}]{ThorntonPhD:1} {Thornton} D., 2013, PhD thesis, Univ. Manchester

\bibitem[\protect\citeauthoryear{Tukey}{1949}]{Tukey:1949:jv} {Tukey} J., 1949,  Biometrics, 5, 2, p.99-114




\bibitem[\protect\citeauthoryear{Way et al.}{2012}]{Way:2012:ut} {Way} M.~J., {Scargle} J.~D., {Ali} K.~M., {Srivastava} A.~N., 2012, Advances in Machine Learning and Data Mining for Astronomy, 1st Edition

\bibitem[\protect\citeauthoryear{Widmer \& Kubat}{1996}]{Widmer:1996:km} {Widmer} G., {Kubat} M., 1996, Machine Learning, 23, 1, p.69

\bibitem[\protect\citeauthoryear{Wilcoxon}{1945}]{Wilcoxon:1945:rs} {Wilcoxon} F., 1945, Biometrics Bulletin, 1, 6, p.80-83




\bibitem[\protect\citeauthoryear{Yang \& Moody}{1999}]{Yang:1999:mo} {Yang} H.~H. \& {Moody}, J., 1999, NIPS, 12, p.687-693

\bibitem[\protect\citeauthoryear{Yang \& Wu}{2006}]{Xindong:2006:yq} {Yang} Q., {Wu} X., 2006, International Journal of Information Technology \& Decision Making, 5, 4, p.597-604


\bibitem[\protect\citeauthoryear{Zhu et al.}{2014}]{Zhu:2014:ab} {Zhu} W.~W. et al., 2014, ApJ, 781, 2


\end{thebibliography}








\bsp	
\label{lastpage}
\end{document}